\PassOptionsToPackage{tablesfirst}{endfloat}
\PassOptionsToPackage{longnamesfirst}{natbib}
\documentclass[useAMS,usenatbib,referee]{biom}
\usepackage{amsmath,amssymb,bm,mathtools}
\usepackage{booktabs,siunitx,enumitem}
\setlist[enumerate]{leftmargin=2em,labelsep=0.5em}

\usepackage[ruled,vlined]{algorithm2e}
\usepackage{graphicx,tikz}
\usetikzlibrary{arrows.meta,positioning,fit,backgrounds}
\usepackage{rotating,multirow,array,color,needspace,etoolbox}
\usepackage{hyperref}
\hypersetup{colorlinks=true,linkcolor=blue,citecolor=blue,urlcolor=blue,filecolor=blue}
\usepackage{biom-compat}
\newcommand{\E}{\mathbb{E}}
\newcommand{\Var}{\mathrm{Var}}

\newcommand{\Normal}{\mathcal{N}}



\newcommand{\condref}[1]{\hyperref[cond:#1]{(#1)}}

\title[RADISH: Robust Historical Control Borrowing]{RADISH: A Robust Covariate-Adaptive Trial Design\\with Historical Control Borrowing}
\author[Zhang et al.]{%
Hengtao Zhang$^{1}$, Yuanke Qu$^{1}$, Mi-Ok Kim$^{2}$, and
Huaqing Jin$^{3,*}$\email{huaqingjin@mail.tsinghua.edu.cn}\\
$^{1}$School of Computer Science and Engineering,\\
Guangdong Ocean University, Guangdong 529500, China\\
$^{2}$Department of Epidemiology and Biostatistics, University of California,\\
San Francisco, California\\
$^{3}$Department of Statistics and Data Science, Tsinghua University,\\
Beijing 100084, China}

\begin{document}
\journal{}
\artmonth{}
\pubyear{}
\shortcites{Neuenschwander2010MAP,Murray2014hierarchical,vanRosmalen2018commensurate,Best2021conflictMAP,Wang2019doublyrobust,Chen2020psborrow,Presanis2013conflict,black2007once,black1996randomised,cummings1998effect,black2020treatment,Nott2020PDC,Banbeta2019MPP,Tsiatis2008CovAdj,Jin2023CAHB,Hobbs2011commensurate,Schmidli2014rMAP,Ibrahim2015powerprior,Viele2014historical,Yang2023SAM,Guo2024AHCC,Bi2023BEATS,Hupf2021semiparMAP,Ghadessi2020roadmap,Lesaffre2024review,Berry2010adaptive,Neuenschwander2020ESS}
\hypersetup{pageanchor=false}
\thispagestyle{empty}
\begin{abstract}
Incorporating historical controls into randomized clinical trials can improve statistical efficiency and alleviate enrollment challenges by reducing the need for concurrent controls, particularly when their recruitment is ethically or logistically challenging. These gains depend on comparability between the historical and concurrent controls. Many dynamic borrowing methods assess compatibility for the overall study populations and apply the same borrowing decision to every participant, even when comparability varies with participant covariates. We propose RADISH, Robust Adaptive Discrepancy-Informed Shrinkage for Historical Borrowing, a covariate-adaptive design that assesses compatibility at the individual participant level and incorporates robust borrowing into sequential randomization and final inference. Under the stated regularity conditions, the RADISH estimator is asymptotically normal and supports asymptotically valid Wald inference. Simulation studies, including semi-synthetic simulation studies based on the two real-world randomized trials, show that RADISH improves efficiency and assigns fewer participants to the control group when the historical and concurrent controls are compatible. As incompatibility increases, RADISH progressively reduces borrowing and approaches a concurrent-only design and analysis.
\end{abstract}
\begin{keywords}
Adaptive design; Covariate-adaptive randomization; Historical borrowing;
Historical--concurrent conflict; Kernel smoothing.
\end{keywords}
\maketitle
\hypersetup{pageanchor=true}

\section{Introduction}
\label{sec:intro}

Randomized clinical trials provide a standard basis for evaluating treatment
efficacy but can require substantial time and resources. Two complementary
strategies can improve their efficiency. Covariate-adaptive randomization balances
prognostic covariates across randomized arms and can improve estimation efficiency
without additional enrollment~\citep{Jiang2018KBCD}. Historical-control borrowing
increases the amount of control information by reusing data from completed studies
\citep{Pocock1976historical,Viele2014historical}. 
This strategy is particularly useful when
concurrent-control recruitment is ethically or logistically difficult
\citep{Ghadessi2020roadmap}. Integrating these strategies makes historical
borrowing part of trial design, so the borrowed information guides both sequential
allocation and final inference according to the degree of compatibility between historical and concurrent controls.

A central consideration in historical borrowing is systematic bias arising
from differences in participant selection and trial conduct between historical and
concurrent populations. Such bias can compromise Type~I error control and the
interpretation of the final analysis
\citep{Viele2014historical,Calderazzo2024robust}. 
This concern has motivated a large literature on Bayesian dynamic borrowing
where borrowing is regulated
through one or more parameters that are
prespecified, estimated empirically, or assigned a hyperprior within a fully
Bayesian framework \citep{Lesaffre2024review}. 
Prominent examples include the power prior, which scales
the historical likelihood~\citep{Ibrahim2015powerprior}; the commensurate prior,
which captures between-trial heterogeneity through a precision hyperparameter
\citep{Hobbs2011commensurate}; and the unit information prior, which measures
borrowed information in single-observation Fisher-information units
\citep{JinYin2021UIP,zhang2023unit,GuYin2024UIDP}. 
Other adaptive approaches,
including self-adapting mixture and elastic priors, use data-driven
compatibility functions to regulate cohort-level borrowing
\citep{Yang2023SAM,Jiang2023elastic}.
A common feature of these approaches is that they assess overall compatibility between historical and concurrent control cohorts and apply the resulting borrowing decision uniformly across participants.

The comparability of historical and concurrent control populations may vary substantially across clinically meaningful patient subgroups, however, even when the studies investigate the same disease and treatment.
In osteoporosis trials, for example, vertebral fracture risk depends on prior fracture history, years since menopause, bone mineral density, and other indicators of underlying skeletal vulnerability \citep{black2007once}.
Although patients can be matched on these observed prognostic characteristics, residual differences in disease severity, prior treatments, and other unmeasured risk factors may lead to varying degrees of comparability between historical and concurrent controls.
Consequently, historical control information may be highly informative for some patient subgroups but less relevant, or potentially misleading, for others.
This heterogeneity motivates a flexible borrowing strategy that accounts for subgroup-specific comparability rather than assuming uniform exchangeability across the study population. 

When compatibility between historical and concurrent controls
differs across participant groups, a global rule may borrow too little for
participants whose historical and concurrent data are compatible and thereby lose
efficiency. It may also borrow too much for participants whose data differ between
the two sources, which can bias treatment-effect estimates.
The Covariate-Adaptive Historical Borrowing (CAHB) design of
\citet{Jin2023CAHB} integrates kernel-based covariate-adaptive design
(KBCD)~\citep{Jiang2018KBCD} with a Bayesian semiparametric model that allows
compatibility to vary across the covariate space.

CAHB uses the estimated historical conditional mean as
a fixed prior center, so the borrowing model does not propagate its sampling
uncertainty. 
As a result, CAHB may borrow too much information from an imprecisely estimated historical mean,
potentially biasing the treatment-effect estimate when historical and concurrent controls differ.
To account for historical estimation uncertainty in local borrowing, we propose
Robust Adaptive Discrepancy-Informed
Shrinkage for Historical Borrowing (RADISH), a
covariate-adaptive clinical trial design that integrates historical
borrowing into sequential randomization and final inference.
RADISH accounts explicitly for the sampling uncertainty of the estimated
historical conditional mean and evaluates its discrepancy from the
concurrent control mean relative to the standard errors of both
estimates. This allows borrowing to depend on both the precision of
historical information and its compatibility with concurrent outcomes.
Motivated by two-sided conflict tail probabilities
\citep{MarshallSpiegelhalter2007PDC,Presanis2013conflict},
RADISH uses a covariate-specific conflict score to discount historical information as local disagreement increases, and the resulting information ratio guides sequential treatment allocation.
At trial completion, a marginal conflict score provides
an additional safeguard for borrowing in the final analysis to ensure its robustness.
For final inference, we use a kernel-standardized estimator for the average treatment effect
in the enrolled cohort, with a variance estimator based on kernel-regression residuals
and linear-smoother weights. This approach provides analytical standard errors
and confidence intervals.
Under the stated regularity conditions, the centered and studentized RADISH
estimator converges to a standard normal distribution, supporting
asymptotically valid Wald inference.

The remainder of the paper is organized as follows.
Section~\ref{sec:method} develops the RADISH trial design,
covering the three-stage design and its large-sample properties.
Section~\ref{sec:sim} compares RADISH with KBCD and CAHB
through simulation studies.
Section~\ref{sec:realdata} presents semi-synthetic simulation studies
based on data from the HORIZON pivotal fracture trial \citep{black2007once} and fracture intervention trial (FIT) \citep{black1996randomised,cummings1998effect}.
Section~\ref{sec:discuss} concludes
and outlines extensions to non-Gaussian endpoints, multi-source
borrowing, and covariate overlap.
Code is publicly available at
\url{https://github.com/BobZhangHT/HistInfoBorrow}.

\section{Methodology: The RADISH Design}
\label{sec:method}

\subsection{Notation and Overview}
\label{subsec:framework}

RADISH is designed for a two-arm randomized clinical trial that sequentially
enrolls participants from a target population. It uses an external historical
control dataset to inform both treatment allocation and final inference.
Let $\mathcal{D}_H = \{(\bm{X}_j^H, Y_j^H)\}_{j=1}^{N_H}$ denote the historical
control data, where $\bm{X}_j^H \in \mathbb{R}^d$ is a baseline covariate vector
and $Y_j^H \in \mathbb{R}$ is a continuous outcome. Let
$\mathcal{D}_n = \{(\bm{X}_i, Y_i, Z_i)\}_{i=1}^n$ denote the concurrent trial
data observed up to interim step~$n$. The treatment indicator is $Z_i = 1$ for
the treatment arm and $Z_i = 0$ for the concurrent control arm. The
superscript~$H$ distinguishes historical observations from concurrent ones
throughout. We write $\mathcal{I}_{z,n} = \{i \le n : Z_i = z\}$ for the index
set of arm $z \in \{0,1\}$ at step~$n$, and $N_z$ for its final size at $n = N$,
so that $N_0 + N_1 = N$.

Let $Y_i(1)$ and $Y_i(0)$ denote the potential outcomes under treatment and
control for participant $i$ in the current trial. We assume consistency, so that
$Y_i=Z_iY_i(1)+(1-Z_i)Y_i(0)$, and write the conditional potential-outcome mean
as $\mu_z(\bm x)=\E\{Y_i(z)\mid \bm X_i=\bm x\}$ for $z\in\{0,1\}$.
Our primary causal estimand is the average treatment effect standardized to
the enrolled trial cohort,
\begin{equation}
\Delta_N
=
\frac1N\sum_{i=1}^N\{\mu_1(\bm X_i)-\mu_0(\bm X_i)\}.
\label{eq:trial_sample_estimand}
\end{equation}
This is a sample-specific estimand defined by the covariate distribution of the
participants enrolled in the trial. Conditional on
$\bm X_1,\ldots,\bm X_N$, $\Delta_N$ is fixed, so the variance describes
randomization and outcome uncertainty for the enrolled covariates.

The design proceeds in three stages.
Stage~I constructs a covariate-specific historical reference and quantifies its uncertainty.
Stage~II compares this reference with the accrued concurrent controls and uses the resulting local borrowing information to guide sequential randomization.
After enrollment is complete, Stage~III combines concurrent and discounted historical control information to estimate $\Delta_N$ and conduct Wald inference. Supplementary Figure~S1 and Algorithm~\ref{alg:radish} summarize this procedure.

For local borrowing, we adopt a Gaussian product kernel $K_{\bm h}$ with bandwidth vector $\bm h\in\mathbb R_+^d$. 
For a covariate profile $\bm x$, define $w_j^H(\bm x)=K_{\bm h}(\bm X_j^H-\bm x)$ and 
$w_i(\bm x)=K_{\bm h}(\bm X_i-\bm x)$,
so that observations with covariates closer to $\bm x$ receive greater weight. The same kernel is used to construct the historical reference, summarize concurrent outcomes, determine local arm information in the randomization rule, and form the final kernel-standardized estimator.
The bandwidth is computed at the planning stage and then fixed across all three stages.
We use Scott's rule of thumb~\citep{scott2015multivariate}, $h_k=N_H^{-1/(d+4)}\hat{\sigma}_{H,k}$, where $\hat{\sigma}_{H,k}$ is the
historical sample standard deviation of covariate $k$.
If $\hat\sigma_{H,k}<10^{-12}$, replace it by one before calculating $h_k$.
Throughout, the similarity weights are evaluated conditionally on the observed covariates, including the historical covariates $\{\bm X_j^H\}_{j=1}^{N_H}$ and, after enrollment, the concurrent covariates $\{\bm X_i\}_{i\le n}$. Thus, conditional on these covariates, the kernel weights are fixed, whereas treatment assignments and outcomes remain random.

Within a local neighborhood of a covariate profile $\bm{x}$ defined by the
kernel weights, we adopt the working Gaussian model
\begin{equation}
Y \mid Z=z,\, \bm{X}\approx\bm{x}
\;\sim\;
\Normal\!\bigl\{\mu_z(\bm{x}),\,\sigma_z^2(\bm{x})\bigr\},
\label{eq:local_working_model_framework}
\end{equation}
where $\mu_z(\bm{x})$ and $\sigma_z^2(\bm{x})$ denote the local conditional
mean and variance in arm $z\in\{0,1\}$. For variance estimation and the
large-sample theory, we further assume homoskedasticity within each concurrent
arm, so that $\sigma_z^2(\bm{x})\equiv\sigma_z^2$ for $z\in\{0,1\}$.

Historical controls provide an external reference for the concurrent control
mean $\mu_0(\bm{x})$. RADISH summarizes the historical data locally by a
reference center $\theta(\bm{x})$ and its estimation variance
$\tau_H^2(\bm{x})=\Var\{\theta(\bm{x})\}$. The quantity
$\tau_H^2(\bm{x})$ characterizes the uncertainty in the estimated historical
reference, whereas $\sigma_0^2$ characterizes the variability of
individual concurrent control outcomes. The construction of
$\theta(\bm{x})$ and $\tau_H^2(\bm{x})$ is described in Stage~I.

\subsection{Stage~I: Construction of the Local Historical Reference}

Stage~I summarizes the historical control data as a covariate-specific
reference for the concurrent control mean. We assume that the historical
outcomes are conditionally independent given their covariates and adopt the
local-constant Gaussian working model
$
Y_j^H \mid \bm X_j^H \approx \bm x
\sim
\Normal\{\mu_H(\bm x),\sigma_H^2(\bm x)\}.
$
Under this model, maximizing the kernel-weighted log-likelihood yields the
local historical mean
\[
\theta(\bm{x})
=
\sum_{j=1}^{N_H} S_{H,j}(\bm{x})Y_j^H,
\qquad
S_{H,j}(\bm{x})
=
\frac{w_j^{H}(\bm{x})}
{\sum_{\ell=1}^{N_H} w_\ell^{H}(\bm{x})}.
\]
Thus, $\theta(\bm{x})$ is the Nadaraya--Watson estimator of
$\mu_H(\bm{x})$.

Under the working model, the reference variance is
$\tau_H^2(\bm x)=\sigma_H^2(\bm x)\sum_jS_{H,j}(\bm x)^2$.
We use the weighted residual variance
$\hat\sigma_H^2(\bm x)=\sum_jS_{H,j}(\bm x)\{Y_j^H-\theta(\bm x)\}^2$
and set $\hat\tau_H^2(\bm x)=\hat\sigma_H^2(\bm x)\sum_jS_{H,j}(\bm x)^2$.
For a fixed local outcome variance, broadly distributed weights give a more
precise historical reference than weights concentrated on a few observations.
After each calculation, local and residual variance estimates are floored at
$\epsilon_v=10^{-12}$ for numerical stability.
Algorithm~S1 (Supplementary Section~S6) constructs the historical reference $\{\theta(\bm x),\hat\tau_H^2(\bm x)\}$ used in subsequent stages.

\subsection{Stage~II: Randomization with Robust Borrowing}

\begin{sloppypar}
For each incoming covariate profile $\bm{x}$, Stage~II discounts historical precision using a local historical--concurrent conflict score motivated by prior-data conflict diagnostics~\citep{Evans2006PDC,Nott2020PDC}. Sequential updates use observed outcomes from previously enrolled participants.
To assess compatibility, we calculate local
concurrent-control summaries using the kernel function, conditional on the
observed control covariates
$\{\bm{X}_i\}_{i \in \mathcal{I}_{0,n}}$.
Define the local kernel mass in arm
 $z\in\{0,1\}$ by $M_{z,n}(\bm x)=\sum_{i\in\mathcal I_{z,n}}w_i(\bm x)$.
 For $i\in\mathcal I_{0,n}$, write
 $S_{0,n,i}(\bm x)=w_i(\bm x)/M_{0,n}(\bm x)$.
 The local summaries are
\begin{align*}
\bar Y_{0,n}(\bm x)&=\sum_{i\in\mathcal I_{0,n}}S_{0,n,i}(\bm x)Y_i,
\qquad
n_{0,n}^{\mathrm{eff}}(\bm x)=\left\{\sum_{i\in\mathcal I_{0,n}}S_{0,n,i}(\bm x)^2\right\}^{-1},\\
\hat\sigma_{0,n}^2(\bm x)&=\sum_{i\in\mathcal I_{0,n}}S_{0,n,i}(\bm x)
\{Y_i-\bar Y_{0,n}(\bm x)\}^2.
\end{align*}
\end{sloppypar}

\begin{sloppypar}
We standardize the difference between $\theta(\bm{x})$ and $\bar{Y}_{0,n}(\bm{x})$ using a reference-predictive standard error that combines historical and concurrent uncertainty.
Specifically, we define
$
\mathrm{SE}_n^2(\bm{x})
=\allowbreak
\hat{\tau}_H^2(\bm{x})
+
{\hat{\sigma}_{0,n}^2(\bm{x})}/
{\widetilde n_{0,n}^{\mathrm{eff}}(\bm{x})}$, 
and 
$
\widetilde n_{0,n}^{\mathrm{eff}}(\bm{x})
=
\max\{n_{0,n}^{\mathrm{eff}}(\bm{x}),n_{\min}\},
$
where $n_{\min}=5$ stabilizes the concurrent sampling-variance term in sparse
neighborhoods. The first term in $\mathrm{SE}_n^2(\bm x)$ reflects uncertainty
in the historical reference, while the second reflects uncertainty in the
concurrent local control mean. We then define the local conflict score as
\begin{equation}
  D_n(\bm{x})
  =
  \left(
  -\log\left[\max\!\left\{
  2\Phi\!\left(
  -\frac{
  \left|\bar Y_{0,n}(\bm{x})-\theta(\bm{x})\right|
  }{
  \mathrm{SE}_n(\bm{x})
  }
  \right),
  \epsilon_c
  \right\}\right]
  -1
  \right)_+,
  \label{eq:local_hcc_standardizer}
  \end{equation}
where $\Phi(\cdot)$ is the standard normal cumulative distribution function, $(u)_+=\max(u,0)$, and $\epsilon_c=10^{-12}$ is a numerical safeguard.
The score increases with standardized disagreement; subtracting one and truncating at zero leaves small discrepancies unpenalized.
\end{sloppypar}

RADISH combines the local conflict score $D_n(\bm{x})$ and the precision of the
historical reference into an allocation-time information ratio. At step $n$,
using only information accrued up to that point, we define
\begin{equation*}
R_n^{A}(\bm x)
=
1+
\frac{\hat\sigma_{0,n}^2(\bm x)}
{\widetilde n_{0,n}^{\mathrm{eff}}(\bm x)\,\mathrm{SE}_n^2(\bm x)}
\exp\{-D_n(\bm x)\}.
\end{equation*}
The information ratio $R_n^A(\bm{x})$ controls borrowing in the randomization
rule, with $R_n^A(\bm{x})=1$ corresponding to no borrowing. Its increment above
one combines historical precision with the conflict discount
$\exp\{-D_n(\bm{x})\}$, reducing borrowing as historical uncertainty or local
conflict increases; see Supplementary Section~S4.

The information ratio $R_n^A(\bm{x})$ is then incorporated into the
randomization rule by augmenting the local control information. Using the local
arm masses $M_{0,n}(\bm{x})$ and $M_{1,n}(\bm{x})$, we assign the next
participant to the treatment arm with probability
\begin{equation}
\phi_n(\bm{x})
=
\frac{\{R_n^{A}(\bm{x})M_{0,n}(\bm{x})\}^2}
{\{R_n^{A}(\bm{x})M_{0,n}(\bm{x})\}^2+\{M_{1,n}(\bm{x})\}^2}.
\label{eq:stage2_allocation}
\end{equation}

This allocation rule extends the kernel-weighted covariate-balancing scheme of
\citet{Jiang2018KBCD} by incorporating the allocation-time borrowing ratio,
following the covariate-adaptive borrowing construction of~\citet{Jin2023CAHB}.
When $R_n^{A}(\bm{x})>1$,
borrowed information enlarges the effective control arm and the rule assigns
fewer participants to the control group. 
As $R_n^{A}(\bm{x})\to1$, it reduces to the
kernel-balanced allocation of~\citet{Jiang2018KBCD}. 
Allocation uses the local discrepancy at the incoming participant's covariate
profile. At trial completion, Stage~III reassesses compatibility using all
concurrent controls and incorporates a marginal conflict score.
Supplementary Algorithm~S2 details sequential allocation.

\subsection{Stage~III: Final Estimation and Inference for the Treatment Effect}
\label{subsec:stage3}

\begin{sloppypar}
At $n=N$, Stage~III estimates $\Delta_N$ using the completed trial data.
Final borrowing weights define a kernel-standardized estimator of $\Delta_N$,
whose outcome-weighted representation supports variance estimation and Wald inference.
To determine the final amount of historical borrowing, we first reassess
historical--concurrent compatibility using the completed trial data. For each
covariate profile $\bm{x}$, the terminal local conflict score $D_N(\bm{x})$ is
obtained from \eqref{eq:local_hcc_standardizer} by setting $n=N$, so that the
comparison uses all concurrent controls observed during the trial.
Although the local conflict score allows borrowing to adapt across the
covariate space, relying on local compatibility alone can lead to unstable
borrowing in finite samples. In our numerical investigations, such instability
can result in excessive historical borrowing and appreciable bias, particularly
when the historical and concurrent control populations differ systematically
in their overall outcome levels. To provide an additional safeguard in the
final analysis, Stage~III therefore supplements the local assessment with a
marginal measure of historical--concurrent conflict.
\end{sloppypar}

Let $\bar Y^C$ and $\bar Y^H$ denote the
overall concurrent- and historical-control means, respectively, with sample
variances $\hat s_C^2$ and $\hat s_H^2$. We define the marginal conflict score
as
$$
D_N^{G}
=
\left(
-\log\left[\max\left\{
2\Phi\left(
-\frac{|\bar Y^H-\bar Y^C|}
{(\hat s_C^2/N_0+\hat s_H^2/N_H)^{1/2}}
\right),
\epsilon_c
\right\}\right]
-1
\right)_+.
$$
The local and marginal conflict scores are then combined through
$
D_N^F(\bm{x})
=
D_N(\bm{x})
+
\{1-\exp[-D_N(\bm{x})]\}D_N^G.
$
The multiplier $1-\exp\{-D_N(\bm{x})\}$ controls the influence of the
marginal conflict score according to the degree of local incompatibility. When
the local score is zero, the marginal score does not affect borrowing at
$\bm{x}$. As the local score increases, the marginal assessment receives
greater weight and provides an additional safeguard against borrowing from an
overall incompatible historical source.

Using the completed concurrent-control cohort, define
$
\widetilde n_{0,N}^{\mathrm{eff}}(\bm{x})
=
\max\{n_{0,N}^{\mathrm{eff}}(\bm{x}),n_{\min}\}
$
and
$
\mathrm{SE}_N^2(\bm{x})
=
\hat{\tau}_H^2(\bm{x})
+
\hat{\sigma}_{0,N}^2(\bm{x})/
\widetilde n_{0,N}^{\mathrm{eff}}(\bm{x}).
$
The final information ratio is
\begin{equation*}
R_N^{F}(\bm{x})
=
1+
\frac{\hat{\sigma}_{0,N}^2(\bm{x})}
{\widetilde n_{0,N}^{\mathrm{eff}}(\bm{x})\,
 \mathrm{SE}_N^2(\bm{x})}
\exp\{-D_N^F(\bm{x})\}.
\end{equation*}
As in Stage~II, the variance ratio accounts for historical precision.
The exponential discount uses the combined local and marginal score $D_N^F(\bm{x})$.
Finally, we convert the information ratio into the borrowing weight
$
W_N^F(\bm{x})
=
(R_N^F(\bm{x})-1)/R_N^F(\bm{x}),
$
which, by construction, is always between $0$ and $1/2$.
A larger $W_N^F(\bm{x})$ corresponds to greater historical borrowing when the
historical reference is precise and compatible with the concurrent controls.
As historical information becomes less precise or less compatible,
$W_N^F(\bm{x})$ decreases toward zero. 

The borrowing weight $W_N^F(\bm{x})$ combines the
concurrent and historical local control estimates to construct the
treatment-effect estimator $\widehat{\Delta}_N$.
Let $(\bm Y_H, \bm Y_0, \bm Y_1)$ denote the outcome vectors for the
historical controls, concurrent controls, and treatment arm, respectively.
For each covariate profile $\bm{x}$, let $\bm S_0(\bm{x})\in\mathbb R^{N_0}$
collect the normalized kernel weights assigned to the concurrent controls, so
that
$
\bm S_0(\bm{x})^\top\bm Y_0
=
\bar Y_{0,N}(\bm{x}).
$
Similarly, let $\bm S_H(\bm{x})\in\mathbb R^{N_H}$ collect the normalized
historical kernel weights introduced in Stage~I, so that
$
\bm S_H(\bm{x})^\top\bm Y_H
=
\theta(\bm{x}).
$
The borrowing-augmented local control mean is
\begin{equation*}
\widehat{\mu}^{\mathrm{aug}}_0(\bm{x})
=
\{1-W_N^F(\bm{x})\}\bm S_0(\bm{x})^\top\bm Y_0
+
W_N^F(\bm{x})\bm S_H(\bm{x})^\top\bm Y_H.
\end{equation*}
The fit reduces to concurrent-only estimation when $W_N^F(\bm{x})=0$.
Since $W_N^F(\bm{x})\leq1/2$, historical controls receive no more weight than
concurrent controls.
Similarly, let $\bm S_1(\bm{x})\in\mathbb R^{N_1}$ denote the row-normalized
kernel weight vector over the concurrent treatment arm.
The local treatment mean is estimated by
$
\widehat{\mu}_1(\bm{x})
=
\bm S_1(\bm{x})^\top\bm Y_1.
$
The treatment-effect estimator averages the fitted
treatment contrast over the enrolled trial covariates:
\begin{equation}
\widehat{\Delta}_N
=
\frac{1}{N}\sum_{i=1}^N
\left\{
\widehat{\mu}_1(\bm X_i)
-
\widehat{\mu}^{\mathrm{aug}}_0(\bm X_i)
\right\}.
\label{eq:stage3_standardized_ate}
\end{equation}
This estimator standardizes the fitted treatment effect to the empirical
covariate distribution of the enrolled trial cohort and is an
empirical-standardization form of the g-computation estimator
\citep{Robins1986gcomputation,Snowden2011gformula}.

For variance estimation, we express
$\widehat{\Delta}_N$ as a weighted sum of the observed outcomes.
Define
$
\widetilde{\bm S}_0(\bm{x})
=
\{1-W_N^F(\bm{x})\}\bm S_0(\bm{x}),
\widetilde{\bm S}_H(\bm{x})
=
W_N^F(\bm{x})\bm S_H(\bm{x}).
$
Define the cumulative coefficient vectors by
$
\bm a_1
=N^{-1}\sum_{i=1}^N\bm S_1(\bm X_i),
\bm a_0
=N^{-1}\sum_{i=1}^N\widetilde{\bm S}_0(\bm X_i),
\bm a_H
=N^{-1}\sum_{i=1}^N\widetilde{\bm S}_H(\bm X_i).
$
Then
\begin{equation}
\widehat{\Delta}_N
=
\bm a_1^\top\bm Y_1
-
\bm a_0^\top\bm Y_0
-
\bm a_H^\top\bm Y_H.
\label{eq:ate_estimator}
\end{equation}
This is algebraically equivalent to \eqref{eq:stage3_standardized_ate}, with
$\bm a_1$, $\bm a_0$, and $\bm a_H$ collecting the respective cumulative treatment,
concurrent-control, and historical-control contributions.
For each source $z\in\{0,1,H\}$, let $\bm S_{zz}$ be its row-normalized kernel
matrix evaluated at its own covariates, including self-weights. We estimate
the source-specific residual variance by
\[
\widehat\sigma_z^2=
\frac{\|\bm Y_z-\bm S_{zz}\bm Y_z\|_2^2}
{\max\{N_z-2\operatorname{tr}(\bm S_{zz})+\|\bm S_{zz}\|_F^2,1\}}.
\]
The denominator is the smoother's residual degrees of freedom.
Here $\|\cdot\|_F$ denotes the Frobenius norm, and $\widehat\sigma_H^2$ is source-wide, distinct from the local
$\widehat\sigma_H^2(\bm x)$ in Stage~I. Treating the coefficients as fixed
under the homoskedastic working model gives the
model-assisted linear-smoother variance
\begin{equation}
\widehat V_{\mathrm{lin}}
=
\widehat{\sigma}_1^2\|\bm a_1\|_2^2
+
\widehat{\sigma}_0^2\|\bm a_0\|_2^2
+
\widehat{\sigma}_H^2\|\bm a_H\|_2^2.
\label{eq:working_variance}
\end{equation}

The borrowing weights entering $\bm a_0$ and $\bm a_H$ are estimated
from the same outcomes used in $\widehat{\Delta}_N$. Treating the realized
weights as fixed may therefore understate finite-sample uncertainty. To keep
the reported variance at least as large as its concurrent-only counterpart, we
also calculate the variance corresponding to the concurrent-only control
estimator. Let
$
\bm a_0^C
=
N^{-1}\sum_{i=1}^N\bm S_0(\bm X_i),
$
and define
$
\widehat V_C
=
\widehat{\sigma}_1^2\|\bm a_1\|_2^2
+
\widehat{\sigma}_0^2\|\bm a_0^C\|_2^2.
$
For computation, $\widehat V_{\mathrm{lin}}$ and $\widehat V_C$ are also floored at $\epsilon_v$.
The reported variance is
$\widehat V_N=\max\{\widehat V_{\mathrm{lin}},\widehat V_C\}$.
Thus, the reported interval is at least as wide as the concurrent-only kernel
interval calculated from the same realized trial data.

The large-sample analysis compares RADISH with a concurrent-only oracle.
Conditional on covariates and randomization draws, its allocation and smoothing
coefficients are outcome-independent. The proof establishes first-order
equivalence of the RADISH and oracle estimators and consistency of the reported
variance, including the concurrent-only maximum.
In Supplementary Section~S5.2, condition (C4) controls the allocation perturbation and requires terminal
borrowing to be first-order negligible, while (C3) and (C6) control smoothing
bias and residual-variance estimation. The following
Studentized limit supports Wald inference.

\begin{theorem}[Asymptotic normality of RADISH estimator]
\label{thm:radish_asymptotic_normality}
Suppose conditions (C1)--(C6) in Supplementary Section~S5.2 hold. Then,
conditional on the
historical data $\mathcal D_H$ and the enrolled covariates
$\bm X_{1:N}$,
\[
\frac{
\widehat{\Delta}_N-\Delta_N
}{
\sqrt{\widehat V_N}
}
\xrightarrow{d}
\Normal(0,1).
\]
\end{theorem}

Supplementary Section~S5 gives the full conditions and proof.
A two-sided $100(1-\alpha)\%$ Wald confidence interval for $\Delta_N$ is
therefore
\begin{equation}
\widehat{\Delta}_N
\pm
z_{1-\alpha/2}\sqrt{\widehat V_N},
\label{eq:wald_ci}
\end{equation}
where $z_{1-\alpha/2}$ is the standard normal $(1-\alpha/2)$ quantile.
The two-sided test rejects $H_0:\Delta_N=0$ at nominal level $\alpha$ when this interval
excludes zero.
Section~\ref{sec:sim} evaluates finite-sample performance.
Algorithm~\ref{alg:radish} summarizes the three stages; Algorithms~S1--S3 in
Supplementary Section~S6 give the detailed computations.

\begin{algorithm}[H]
\SetAlgoLined
\SetKwInput{KwIn}{Input}
\SetKwInput{KwOut}{Output}
\KwIn{Historical controls $\mathcal D_H$, trial size $N$, $n_{\mathrm{init}}=20$, and $\alpha=0.05$.}
\KwOut{Assignments $\{Z_i\}_{i=1}^N$, estimate $\widehat\Delta_N$, and Wald interval \eqref{eq:wald_ci}.}
Fix $\bm h$ and construct $\{\theta(\bm x),\hat\tau_H^2(\bm x)\}$ as in Stage~I\;
Randomize the first $n_{\mathrm{init}}$ participants with probability $1/2$ and observe their outcomes\;
\For{$n=n_{\mathrm{init}},\ldots,N-1$}{
Observe $\bm X_{n+1}$; compute local control summaries, arm masses, $D_n$, and $R_n^A$\;
Compute $\phi_n$ by \eqref{eq:stage2_allocation}; draw $Z_{n+1}$, observe $Y_{n+1}$, and append the record\;
}
Using all concurrent controls, recompute $D_N(\bm X_i)$ and the marginal score $D_N^G$; obtain $D_N^F(\bm X_i)$, $R_N^F(\bm X_i)$, and $W_N^F(\bm X_i)$ as in Stage~III\;
Form the fitted means and estimate $\widehat\Delta_N$ by \eqref{eq:stage3_standardized_ate}\;
Compute the residual variances, $\widehat V_{\mathrm{lin}}$ in \eqref{eq:working_variance}, and $\widehat V_C$; set $\widehat V_N=\max\{\widehat V_{\mathrm{lin}},\widehat V_C\}$\;
Compute $\widehat{\mathrm{se}}_N=\sqrt{\widehat V_N}$ and return $[\widehat\Delta_N\pm z_{1-\alpha/2}\widehat{\mathrm{se}}_N]$\;
\caption{The three-stage RADISH design.}
\label{alg:radish}
\end{algorithm}

\section{Simulation Studies}
\label{sec:sim}

\subsection{Simulation Settings and Data-Generating Process}
\label{subsec:sim_setup}

We evaluated sequential allocation and final inference using external historical
controls and a sequentially enrolled two-arm trial, with fixed sample sizes
$N_H = N = 200$ and $M=1000$ independent replications per scenario.
After $n_{\mathrm{init}}=20$ equal-probability burn-in assignments, the adaptive
allocation rule was activated. We used the Gaussian product kernel
and fixed historical-data bandwidth specified in Section~\ref{subsec:framework}, with $d=2$
and Scott's rule~\citep{scott2015multivariate}.

Each participant was assigned $\bm X=(X_1,X_2)^\top$, with
$X_1\sim\mathrm{Bernoulli}(0.5)$ and $X_2\sim\mathcal N(0,1)$. The concurrent
outcome was $Y=\mu_0^{(C)}(\bm X)+\Delta Z+\varepsilon_Z$, where
$\mu_0^{(C)}(\bm X)=\beta_0+\beta_1X_1+\beta_2X_2+\beta_3X_1X_2$. We set
$\beta_0=0$, $\beta_1=\beta_2=0.5$, and $\beta_3=0.3$, with
$\varepsilon_0\sim\mathcal N(0,1.0^2)$ and
$\varepsilon_1\sim\mathcal N(0,1.2^2)$. 
Historical outcomes followed
$Y^H=\mu_0^{(C)}(\bm X)+b(X_1)+\varepsilon_H$, where
$\varepsilon_H\sim\mathcal N(0,\sigma_H^2)$ and
$b(x_1)=b(0)\mathbf 1\{x_1=0\}+b(1)\mathbf 1\{x_1=1\}$. Thus
$\mu_H(\bm x)=\mu_0^{(C)}(\bm x)+b(x_1)$.
In the simulation,
the historical outcome variance is homoskedastic, so that
$\sigma_H(\bm x)\equiv\sigma_H$ for all $\bm x$.
In every nonzero-conflict
scenario, the historical surface differs from the concurrent surface in both
$X_1$ strata, with a larger shift when $X_1=1$. Consequently, the overall mean shift
$\{b(0)+b(1)\}/2$ can conceal differences between the two strata.

\begin{sloppypar}
We considered $\Delta\in\{0,0.5\}$. Because the treatment effect is constant,
$\mu_1(\bm x)-\mu_0(\bm x)=\Delta$ for every $\bm x$, and hence
$\Delta_N=\Delta$ in every trial. 
The primary study uses a
$4\times2$ conflict-by-precision design with
$(b(0),b(1))\in\{(0,0),(0.04,0.12),(0.25,0.75),(1.5,2.5)\}$ and
$\sigma_H\in\{0.5,3.0\}$, defining scenarios S1--S8. S1/S2 are exactly
compatible, S3/S4 have small covariate-dependent conflict, S5/S6 have moderate
conflict, and S7/S8 have severe conflict. 
All nonzero-conflict scenarios are covariate dependent, with different
historical shifts for $X_1=0$ and $X_1=1$. Within each conflict level, we
consider a precise historical source with $\sigma_H=0.5$ and a less precise
historical source with $\sigma_H=3.0$.
\end{sloppypar}

We compared RADISH with kernel-based covariate-adaptive design (KBCD) \citep{Jiang2018KBCD} and covariate-adaptive historical borrowing (CAHB) \citep{Jin2023CAHB}.
KBCD fixes $R_n^{\mathrm{KBCD}}(\bm x) = 1$ and uses only
concurrent outcomes. 
Like the other methods, it uses the prespecified planning bandwidth computed from historical covariates.
 It therefore serves as the concurrent-only inference reference.  
CAHB uses one
fitted local model to calculate posterior variances with its borrowing term
disabled and enabled, denoted by
$\mathrm{Var}_{\mathrm{ref}}\{\mu_0(\bm x)\}$ and
$\mathrm{Var}_{\mathrm{borrow}}\{\mu_0(\bm x)\}$, respectively, and sets
$R_n^{\mathrm{CAHB}}(\bm x)=\mathrm{Var}_{\mathrm{ref}}\{\mu_0(\bm x)\}/
\mathrm{Var}_{\mathrm{borrow}}\{\mu_0(\bm x)\}$.
The implemented comparator uses $\gamma=0.25$ and $\lambda_2(n)=300\max\{\log n,1\}$.
All three methods use the allocation rule in \eqref{eq:stage2_allocation}. The CAHB comparator uses the common terminal kernel estimator and residual-based variance calculation, so this comparison evaluates its borrowing and allocation rules within the shared final-analysis framework. RADISH additionally applies the concurrent-only variance floor described in Section~\ref{subsec:stage3}.

For each scenario and method, we evaluated the empirical bias, the
root mean squared error (RMSE), the Type~I error (rejection rate under $\Delta = 0$), and calibrated
power under $\Delta = 0.5$. 
Calibrated power was calculated using the empirical 95th percentile
of the absolute null Wald statistic as its
critical value~\citep{lloyd2005estimating}. 
We also evaluated the empirical
confidence-interval coverage probability, the average
confidence-interval width, and the allocation proportion to the
control arm, $N_0/N$. 
To describe how borrowing changes along the bias and
precision axes, we also recorded three borrowing measures:
the mean borrowing weight $\overline W_N^F = N^{-1}\sum_i W_{N,i}^F$, with
$W_{N,i}^F=(R_{N,i}^F-1)/R_{N,i}^F$,
the mean local information ratio
$\overline R_N^F = N^{-1}\sum_i R_{N,i}^F$, and the mean combined
conflict score
$\overline D_N^F = N^{-1}\sum_i D_N^F(\bm X_i)$. 
Each measure was averaged across enrolled participants and replications.

\subsection{Hypothesis Testing and Estimation Performance}
\label{subsec:sim_results_testing}

\begin{figure}[!ht]
\centering
\includegraphics[width=\textwidth]{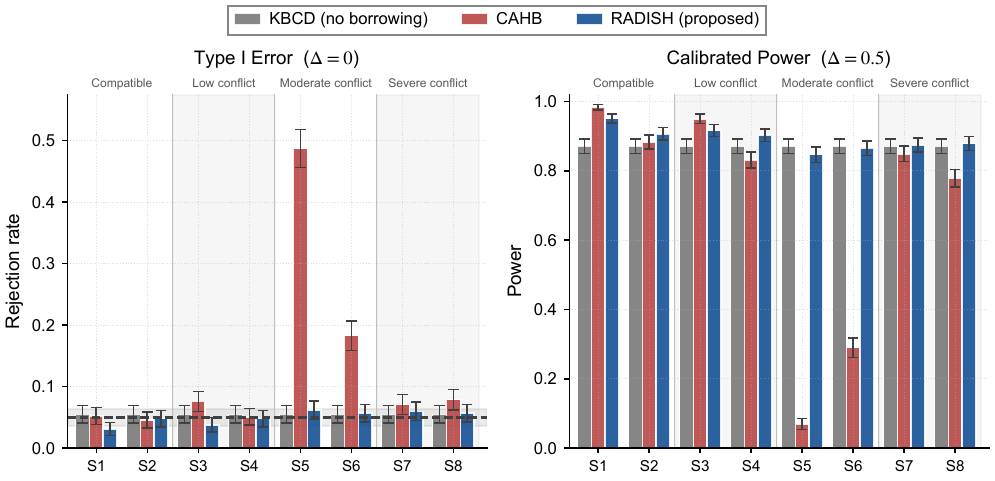}
\caption{Hypothesis-testing performance across the eight simulation
scenarios at $M = 1000$ replications. Left panel: empirical Type~I
error rate under the null hypothesis ($\Delta = 0$); the
horizontal dashed line marks the nominal $\alpha = 0.05$ level, and the
shaded band is the normal-approximation $95\%$ confidence interval for a correctly
calibrated test at this $M$.
The right panel shows calibrated power under the alternative
hypothesis ($\Delta = 0.5$), using the empirical
$0.95$ null quantile as the critical value. 
Error bars are pointwise
normal-approximation $95\%$ confidence intervals. 
Scenarios S1--S8 use
$(b(0),b(1))=(0,0),(0.04,0.12),(0.25,0.75),(1.5,2.5)$, each at
$\sigma_H=0.5$ and $3.0$, with $N_H=N=200$. }
\label{fig:testing}
\end{figure}

Figure~\ref{fig:testing} shows hypothesis-testing performance and 
Table~\ref{tab:metrics_power} reports estimation, interval, allocation, and
borrowing results under the alternative. 
The corresponding null results are given in Supplementary Table~S1.
 With $M=1000$ replications, the normal-approximation 95\% Monte Carlo
 reference interval around a nominal Type~I error of $0.05$ is $[0.036,0.064]$.
RADISH's Type~I errors range from $0.031$ to $0.062$ across the eight scenarios.
Most of the Type~I errors are within the reference interval
apart from the conservative S1 result ($0.031$). 
CAHB is well calibrated under
compatibility but records $0.487/0.183$ under moderate conflict with high and low precision historical information, respectively.
This inflation may reflect CAHB's assessment of discrepancy after shrinking
the concurrent control mean toward the historical mean, which can make
moderate conflict appear small enough to retain substantial borrowing.
The concurrent-only KBCD rejection rate is $0.055$ throughout since it does not borrow from the historical data.

\begin{table}[!ht]
\centering
\caption{Estimation, interval, allocation, and borrowing performance under
$\Delta=0.5$, based on $M=1000$ replications with $N_H=N=200$.
Scenarios S1--S8 combine standardized historical shifts
$(b(0),b(1))=(0,0),(0.04,0.12),(0.25,0.75),(1.5,2.5)$ with
$\sigma_H=0.5,3.0$, respectively. RMSE is root mean squared error; Width is
mean 95\% interval width; Alloc. Proportion is the proportion assigned to the control group;
$\overline W_N^F$ is the mean Stage~III borrowing weight.
}
\label{tab:metrics_power}
\sisetup{round-mode=places, round-precision=3, table-format=-1.3}
\resizebox{0.92\textwidth}{!}{%
\begin{tabular}{l l S[table-format=-1.3] S[table-format=1.4, round-precision=4] S[table-format=1.3] S[table-format=1.4, round-precision=4] S[table-format=1.3] S[table-format=1.3]}
\Hline
{Scenario} & {Method} & {Bias} & {RMSE} & {Coverage} & {Width} & {Alloc.\ Proportion} & {$\overline W_N^F$} \\
\hline
\multirow{3}{*}{S1: Compatible, $\sigma_H\!=\!0.5$}
  & KBCD   &  0.002 & 0.1507 & 0.969 & 0.6167 & 0.500 & 0.000 \\
  & CAHB   &  0.003 & 0.1175 & 0.961 & 0.4897 & 0.407 & 0.511 \\
  & RADISH &  0.004 & 0.1362 & 0.967 & 0.6175 & 0.420 & 0.382 \\
\hline
\multirow{3}{*}{S2: Compatible, $\sigma_H\!=\!3.0$}
  & KBCD   &  0.002 & 0.1507 & 0.969 & 0.6167 & 0.500 & 0.000 \\
  & CAHB   &  0.006 & 0.1557 & 0.963 & 0.6524 & 0.414 & 0.478 \\
  & RADISH &  0.003 & 0.1495 & 0.963 & 0.6147 & 0.473 & 0.127 \\
\hline
\multirow{3}{*}{S3: Small, $\sigma_H\!=\!0.5$}
  & KBCD   &  0.002 & 0.1507 & 0.969 & 0.6167 & 0.500 & 0.000 \\
  & CAHB   & -0.036 & 0.1232 & 0.952 & 0.4898 & 0.407 & 0.511 \\
  & RADISH & -0.016 & 0.1374 & 0.966 & 0.6176 & 0.421 & 0.370 \\
\hline
\multirow{3}{*}{S4: Small, $\sigma_H\!=\!3.0$}
  & KBCD   &  0.002 & 0.1507 & 0.969 & 0.6167 & 0.500 & 0.000 \\
  & CAHB   & -0.026 & 0.1585 & 0.961 & 0.6520 & 0.414 & 0.476 \\
  & RADISH & -0.002 & 0.1484 & 0.962 & 0.6148 & 0.473 & 0.126 \\
\hline
\multirow{3}{*}{S5: Moderate, $\sigma_H\!=\!0.5$}
  & KBCD   &  0.002 & 0.1507 & 0.969 & 0.6167 & 0.500 & 0.000 \\
  & CAHB   & -0.225 & 0.2638 & 0.549 & 0.4962 & 0.412 & 0.487 \\
  & RADISH & -0.021 & 0.1531 & 0.960 & 0.6160 & 0.456 & 0.138 \\
\hline
\multirow{3}{*}{S6: Moderate, $\sigma_H\!=\!3.0$}
  & KBCD   &  0.002 & 0.1507 & 0.969 & 0.6167 & 0.500 & 0.000 \\
  & CAHB   & -0.146 & 0.2244 & 0.846 & 0.6448 & 0.424 & 0.416 \\
  & RADISH & -0.011 & 0.1510 & 0.956 & 0.6151 & 0.478 & 0.088 \\
\hline
\multirow{3}{*}{S7: Severe, $\sigma_H\!=\!0.5$}
  & KBCD   &  0.002 & 0.1507 & 0.969 & 0.6167 & 0.500 & 0.000 \\
  & CAHB   & -0.005 & 0.1621 & 0.940 & 0.6152 & 0.497 & 0.005 \\
  & RADISH &  0.001 & 0.1503 & 0.974 & 0.6166 & 0.499 & 0.001 \\
\hline
\multirow{3}{*}{S8: Severe, $\sigma_H\!=\!3.0$}
  & KBCD   &  0.002 & 0.1507 & 0.969 & 0.6167 & 0.500 & 0.000 \\
  & CAHB   & -0.025 & 0.1712 & 0.927 & 0.6167 & 0.492 & 0.028 \\
  & RADISH & -0.001 & 0.1510 & 0.973 & 0.6163 & 0.497 & 0.004 \\
\hline
\end{tabular}}
\end{table}

RADISH attains calibrated power $0.951/0.907$ under compatibility and
$0.916/0.903$ under small conflict, compared with $0.871$ for KBCD. In the
moderate-conflict scenarios, RADISH records $0.847/0.865$, whereas CAHB falls to
$0.070/0.290$ as its negative bias offsets the treatment effect. Under severe
conflict, RADISH records $0.875/0.879$, close to KBCD, while its mean borrowing
weights fall to $0.001/0.004$. 

With precise historical information, RADISH reduces RMSE from $0.1507$ for
KBCD to $0.1362$ under compatibility and $0.1374$ under small conflict
(Table~\ref{tab:metrics_power}). These gains diminish with lower historical
precision. Under moderate conflict, RADISH retains small biases and RMSEs
near KBCD's, whereas CAHB exhibits substantially larger errors. Under severe
conflict, RADISH approaches KBCD as borrowing vanishes. Coverage under the
alternative ranges from $0.956$ to $0.974$, reflecting the concurrent-only
variance safeguard while preserving estimation gains from informative borrowing.

\subsection{Allocation Behavior and Borrowing Diagnostics}
\label{subsec:sim_results_alloc_diag}

\begin{figure}[!ht]
\centering
\includegraphics[width=\textwidth]{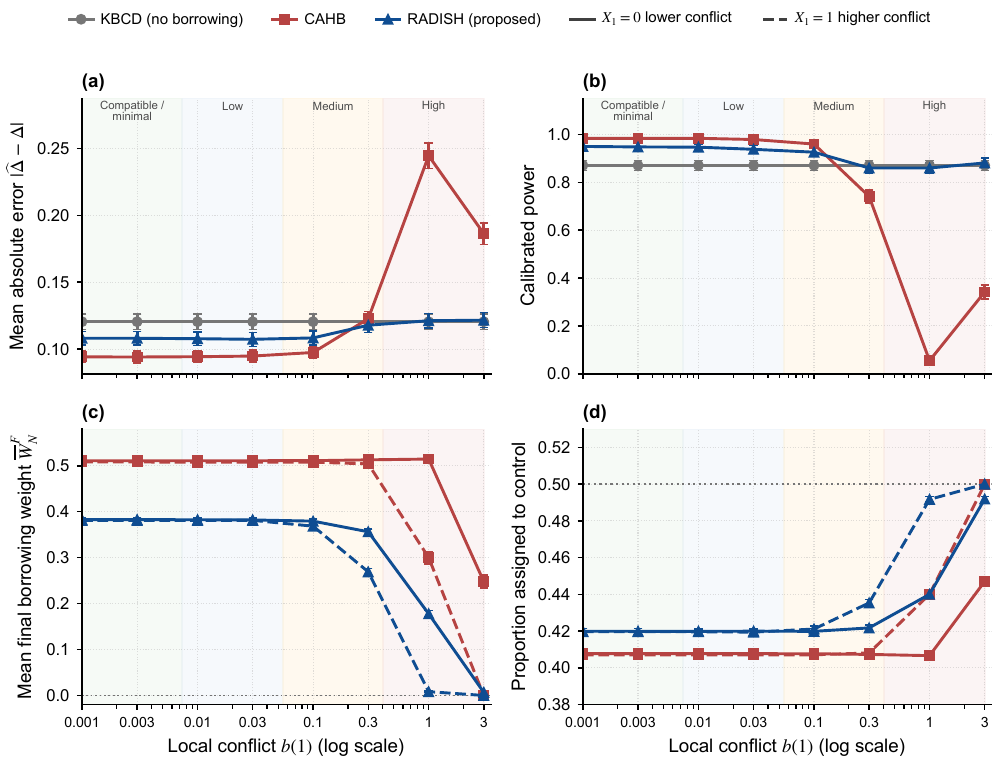}
\caption{Local borrowing performance under $\sigma_H=0.5$, $\Delta=0.5$ and $N_H=N=200$ based on $M=1000$ replications.
 The path uses $b(1)\in\{0.001,\allowbreak0.003,\allowbreak0.01,\allowbreak0.03,\allowbreak0.10,\allowbreak0.30,\allowbreak1,\allowbreak3\}$ and
$b(0)=b(1)/3$, so every
nonzero conflict depends on $X_1$ and is stronger in the $X_1=1$ stratum. Panels show
(a) mean absolute error, (b) calibrated power using an empirical
Type~I error of $0.05$ within each scenario and method, (c) the mean terminal
borrowing weight by stratum, and (d) the proportion assigned to the control group by
stratum. Solid and dashed curves denote the lower- and higher-conflict strata,
respectively. 
}
\label{fig:diag}
\end{figure}

Table~\ref{tab:metrics_power} and Supplementary Figure~S2 summarize allocation, while
Figure~\ref{fig:diag} shows local borrowing. KBCD's control allocation remains
near $0.50$. With precise, compatible historical information (S1), CAHB/RADISH
allocate $0.407/0.420$ to control, with similar proportions under small conflict
(S3). RADISH moves toward equal allocation as conflict increases or historical
precision decreases. Under moderate conflict (S5), its proportion is $0.456$
versus $0.412$ for CAHB; under severe conflict (S7/S8), it is nearly $0.50$.

Figure~\ref{fig:diag} further illustrates the local borrowing mechanism along
an eight-point conflict path with fixed historical precision
($\sigma_H=0.5$), where
$b(1)\in\{0.001,\allowbreak0.003,\allowbreak0.01,\allowbreak0.03,\allowbreak0.10,\allowbreak0.30,\allowbreak1,\allowbreak3\}$ and
$b(0)=b(1)/3$.
Along this path, RADISH retains its estimation and power gains under small
conflict, while the borrowing weights for the two $X_1$ strata gradually
separate as the covariate-specific discrepancies become more pronounced.
As conflict increases further, both borrowing weights approach zero and the
performance of RADISH converges toward that of KBCD. CAHB remains efficient
near compatibility, but its estimation accuracy and
calibrated power deteriorate substantially over the moderate-conflict range.
Supplementary Figure~S3 further examines how borrowing responds
to the mean shift and variation in subgroup-specific shifts.
The sensitivity analysis with a larger historical sample size,
$N_H=400$, shows a similar qualitative pattern (Supplementary Figure~S4), with somewhat greater efficiency
gains under compatibility and small conflict while retaining the same
conflict-adaptive pattern. The HORIZON--FIT study below evaluates this
borrowing behavior using covariate distributions and an outcome model calibrated
to clinical trial data.

\section{Semi-Synthetic Simulation Studies Based on HORIZON--FIT Data}
\label{sec:realdata}

\subsection{HORIZON--FIT Data and Outcome Model}
\label{subsec:realdata_data_model}

We evaluated borrowing under overall and subgroup-specific mean shifts using
semi-synthetic data from HORIZON~\citep{black2007once} and
FIT~\citep{black1996randomised,cummings1998effect}.
We used HORIZON data to estimate the covariate distribution and fit
the outcome regression model for the concurrent trial, and FIT data
to estimate the covariate distribution for the historical controls.
Both were placebo-controlled randomized
trials involving postmenopausal women at elevated fracture risk.
We used total hip bone mineral density (BMD) at 24 months as the outcome.
This outcome was chosen because previous research supports using
treatment-related percentage changes in total hip BMD as a surrogate
for reductions in vertebral fracture risk~\citep{black2020treatment}.

We restricted the analysis to White women aged at least 80 years
in HORIZON and at least 78 years in FIT. After excluding participants
with missing values, the cohorts comprised
620 HORIZON participants and 388 FIT participants.
We included four prognostic covariates:
fall history ($X_1$, $X_1=1$ if there is any fall history, otherwise 0),
vertebral fracture history ($X_2$, $X_2 = 1$ if there is any vertebral fracture history, otherwise 0),
years since menopause ($X_3$) and the baseline hip BMD ($X_4$).

We standardized 24-month hip BMD using the mean and standard deviation from the HORIZON trial 
and modeled the standardized outcome as
\begin{equation}
\label{eq:realmodel}
\widetilde{Y}
\;=\;
\beta_0(X_1, X_2)
+ \beta_1(X_1, X_2)\,\widetilde{X}_3
+ \beta_2(X_1, X_2)\,\widetilde{X}_4
+ \delta\,Z
+ \varepsilon_C,
\qquad
\varepsilon_C \sim \mathcal{N}(0, \sigma_C^2).
\end{equation}
The continuous covariates
$\widetilde{X}_3$ and $\widetilde{X}_4$ were also standardized using the corresponding
means and standard deviations from the HORIZON trial.
The four subgroups were indexed by
$(X_1,X_2)\in\{(0,0),(1,0),(0,1),(1,1)\}$. 

We used ordinary least squares within each HORIZON subgroup to
estimate the subgroup-specific intercepts and covariate coefficients
in \eqref{eq:realmodel}.
We also estimated the conditional distributions of
$(X_3,X_4)\mid(X_1,X_2)$ separately in HORIZON and FIT using Gaussian
kernel density estimation, with bandwidths selected using Scott's
rule of thumb~\citep{scott2015multivariate}.
For each concurrent participant, we first sampled $(X_1,X_2)$ using the empirical subgroup
proportions in HORIZON, approximately
$(0.198, 0.082, 0.492, 0.227)$, and then sampled $(X_3,X_4)$ from
the corresponding estimated conditional distribution.
Given these covariates and treatment assignment, we generated
outcomes from \eqref{eq:realmodel}, with $\delta=0$ under the null
hypothesis, and $\delta=0.14$ under the alternative. 
The residual standard deviation was set to $\sigma_C=\sqrt{0.11}$, which was estimated from HORIZON data.
Historical covariates were sampled analogously from the estimated FIT distribution.
We then generated outcomes from \eqref{eq:realmodel} with $Z=0$
and an additional mean shift $b(X_1)$.
The same residual standard deviation was used for historical outcomes, i.e., $\sigma_H=\sigma_C$.
The shift $b(X_1)$ was specified according to the conflict scenarios
described in the next subsection.
Table~\ref{tab:realsetup} summarizes the simulation parameters.

\begin{table}[!ht]
  \centering
  \caption{
  Simulation settings and estimated model parameters for the
  semi-synthetic simulation studies.
  Subgroup-specific values correspond to
  $(X_1,X_2)=(0,0)$, $(1,0)$, $(0,1)$, and $(1,1)$, respectively.
  }
  \label{tab:realsetup}
  \resizebox{\textwidth}{!}{%
  \begin{tabular}{@{}lll@{}}
  \Hline
  Parameter & Symbol & Value \\
  \hline
  \multicolumn{3}{@{}l}{\textit{Parameters estimated from the data}} \\
  Subgroup proportions in FIT
    & ---
    & $(0.346,\,0.144,\,0.366,\,0.144)$ \\
  Subgroup proportions in HORIZON
    & ---
    & $(0.198,\,0.082,\,0.492,\,0.227)$ \\
  Subgroup intercepts from HORIZON
    & $\beta_0(X_1,X_2)$
    & $(-0.203,\,-0.238,\,-0.150,\,-0.176)$ \\
  Coefficient for $\widetilde X_3$ from HORIZON
    & $\beta_1(X_1,X_2)$
    & $(0.003,\,-0.053,\,0.010,\,0.001)$ \\
  Coefficient for $\widetilde X_4$ from HORIZON
    & $\beta_2(X_1,X_2)$
    & $(0.925,\,0.963,\,0.939,\,0.911)$ \\
  Proportion with a history of falls in HORIZON
    & $p_1$
    & $192/620\approx0.310$ \\
  \hline
  \multicolumn{3}{@{}l}{\textit{Simulation settings}} \\
  Concurrent sample size
    & $N$ & $200$ \\
  Historical sample size
    & $N_H$ & $350$\\
  Initial equal-randomization sample size
    & $n_{\mathrm{init}}$ & $20$ \\
  Replications per method and scenario
    & $M$ & $1000$  \\
  Treatment effect
    & $\delta$ & $0$ (null); $0.14$ (alternative) \\
  Concurrent residual standard deviation
    & $\sigma_C$ & $\sqrt{0.11}$ \\
  Historical residual standard deviation
    & $\sigma_H$ & $\sigma_C$ \\
  \hline
  \multicolumn{3}{@{}l}{\textit{First set of scenarios}} \\
  Standardized mean shift
    & $b(X_1)/\sigma_C$
    & $g+\ell(X_1-p_1)/\sqrt{p_1(1-p_1)}$ \\
  Average standardized shift
    & $g$ & $\{0,\,2\}$ \\
  Standard deviation of standardized shifts
    & $\ell$ & $\{0.1,\,\sqrt{(1-p_1)/p_1}\}$ \\
  \hline
  \multicolumn{3}{@{}l}{\textit{Second set of scenarios}} \\
  Standardized subgroup shifts
    & $(b(0),b(1))/\sigma_C$
    & $4\xi\bigl(-1,(1-p_1)/p_1\bigr)$ \\
  Shift magnitude
    & $\xi$
    & $\{0,\,0.003162,\,0.01,\,0.031623,\,0.1,\,0.316228,\,1\}$ \\
  \hline
  \end{tabular}%
  }
\end{table}

\subsection{Simulation Settings and Evaluation}
\label{subsec:realdata_scenarios}

We considered two sets of scenarios. The first varied the overall mean shift and its variation
across fall-history subgroups ($X_1=0$ and $X_1=1$). The second increased
subgroup differences while holding the overall mean shift at zero.
All overall shifts were averaged over the estimated HORIZON covariate distribution.

For the first set of scenarios, let $p_1=192/620\approx0.310$
denote the proportion of HORIZON participants with a history
of falls. We specified the historical mean shift as
$$
\frac{b(X_1)}{\sigma_C}
=
g+\ell\frac{X_1-p_1}{\sqrt{p_1(1-p_1)}}.
$$
Under the HORIZON covariate distribution, the standardized shift
$b(X_1)/\sigma_C$ has mean $g$ and standard deviation $\ell$.
Thus, $g$ is the HORIZON-averaged standardized conditional mean shift, and $\ell$ measures its variation across fall-history subgroups. Because covariate distributions differ, the unadjusted marginal control-mean difference, on the modeled outcome scale, need not equal $\sigma_C g$.

We considered all four combinations of $g\in\{0,2\}$ and
$\ell\in\{0.1,\sqrt{(1-p_1)/p_1}\}$, where
$\sqrt{(1-p_1)/p_1}\approx1.493$.
The two values of $g$ correspond to no mean shift and a mean 
shift of two residual standard deviations.
The two values of $\ell$ represent small and large variation in subgroup-specific mean shifts.
For example, when $g=0$ and $\ell=\sqrt{(1-p_1)/p_1}$, the scenario has substantial
subgroup-specific shifts that are cancelled when averaged over the
HORIZON population. This scenario examines whether borrowing
methods detect subgroup-specific discrepancies when the mean
discrepancy is zero.

The second set extended the range of subgroup-specific shifts while preserving
a zero overall mean shift. We set
$$
\frac{(b(0),b(1))}{\sigma_C}
=
4\xi\left(-1,\frac{1-p_1}{p_1}\right),
$$
with
$\xi\in\{0,0.003162,0.01,0.031623,0.1,0.316228,1\}$.
At $\xi=0$, the historical and concurrent conditional control
means coincide.
As $\xi$ increases, the shifts in the two fall-history subgroups
increase in magnitude with opposite signs.
Their average under the HORIZON covariate distribution remains
zero. 
At $\xi=1$, the standardized shifts are approximately
$(-4.00,8.91)$.

For each scenario,
we generated $N_H=350$ historical controls
and $N=200$ concurrent participants.
We conducted $M=1000$ simulation replications for each method
under both the null and alternative hypotheses.
Each trial used $n_{\mathrm{init}}=20$ equal-probability burn-in assignments
before method-specific allocation.
Because the treatment effect was constant across covariates,
$\Delta_N=\delta$ in every replication.
Table~\ref{tab:realsetup} summarizes the simulation settings.

We compared RADISH, KBCD, and CAHB as in Section~\ref{sec:sim}.
RADISH used the same parameter settings as in that section.
For CAHB, we set $\gamma=\sqrt{3}$ and
$\lambda_2(n)=300\max\{\log n,1\}$.
We evaluated the proportion assigned to the control group, 
the mean
borrowing weight $\overline W_N^F$, estimation bias, mean absolute
error (MAE), Type~I error, and calibrated power.
Mean absolute error was calculated as
$M^{-1}\sum_{r=1}^M|\widehat\Delta_{N,r}-\delta|$,
where $\widehat\Delta_{N,r}$ is the treatment-effect estimate
in replication $r$.
Calibrated power used the empirical null critical value
estimated separately for each method and scenario, as described
in Section~\ref{sec:sim}.
The resulting power comparisons use a common empirical null rejection rate
for each scenario.

\subsection{Results under Semi-Synthetic Simulation Studies}
\label{subsec:realdata_results}

\begin{figure}[!ht]
\centering
\includegraphics[width=\textwidth]{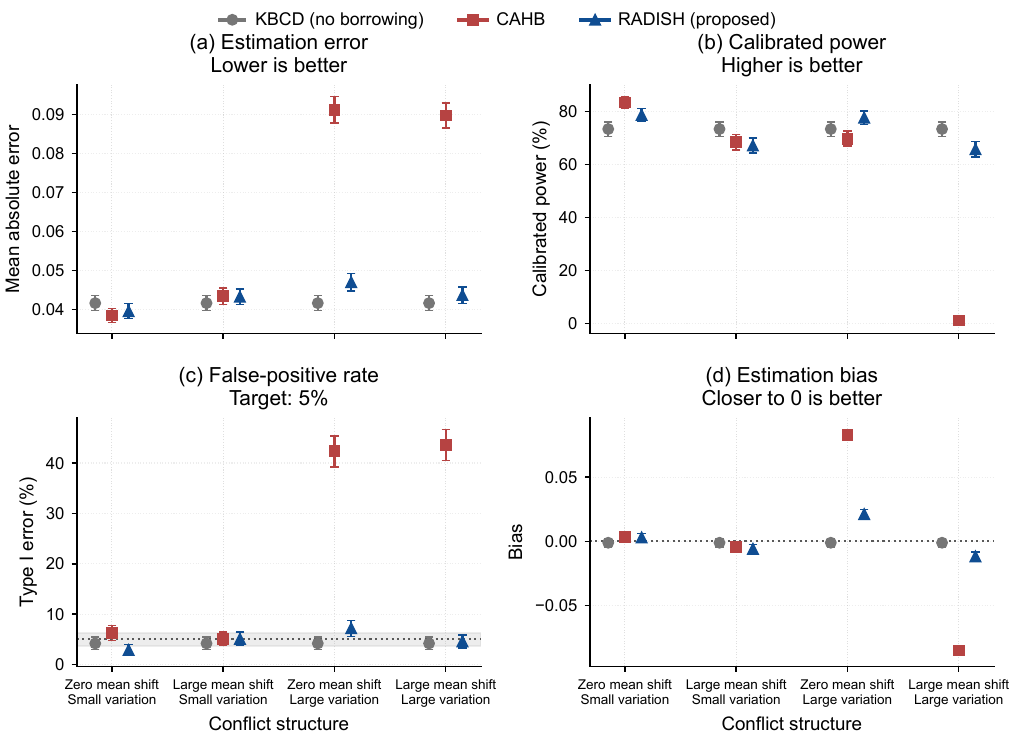}
\caption{
  Performance across four HORIZON--FIT simulation scenarios, with
$N=200$, $N_H=350$, and $1000$ replications per scenario.
The HORIZON-averaged standardized conditional shift is labeled zero ($g=0$) or large ($g=2$).
Variation in subgroup-specific shifts is labeled small
($\ell=0.1$) or large ($\ell=1.493$).
Panels show (a) mean absolute error, (b) calibrated power,
(c) empirical Type~I error, and (d) estimation bias.
Panels (a), (b), and (d) use $\delta=0.14$, and panel (c) uses
$\delta=0$.
The dotted lines indicate Type~I error $0.05$ and zero bias.
Error bars show pointwise normal-approximation $95\%$ confidence intervals. The gray band in panel (c) is the corresponding Monte Carlo reference at nominal Type~I error. Calibrated power uses method- and scenario-specific empirical null critical values.
}
\label{fig:realF1}
\end{figure}

Figure~\ref{fig:realF1} summarizes the results under the first set of scenarios.
Compared with concurrent-only KBCD, RADISH improves estimation
accuracy and calibrated power when the historical
and concurrent controls are similar.
With zero mean shift ($g=0$) and small variation in subgroup-specific
shifts ($\ell=0.1$), RADISH reduces MAE from $0.042$ to $0.040$
and increases calibrated power from $0.734$ to $0.787$.
The empirical Type~I error is $0.029$.
In the remaining scenarios, RADISH is overall comparable to KBCD, 
with similar MAE, Type~I error, and calibrated power.
Nevertheless, when the mean shift is zero and variation in subgroup-specific shifts 
is large ($\ell\approx1.493$), RADISH shows modest Type~I error inflation,
with a rejection rate of $0.072$.

Compared with CAHB, RADISH borrows more conservatively and is
more robust when the shifts vary substantially across subgroups.
The two methods perform similarly when variation in
subgroup-specific shifts is small.
With large variation, however, CAHB shows substantial Type~I
error inflation: its rejection rates are $0.423$ for $g=0$
and $0.436$ for $g=2$.
The corresponding RADISH rates are $0.072$ and $0.046$.
RADISH also has lower MAE in both scenarios, with values of
$0.047$ and $0.044$, compared with $0.091$ and $0.090$ for CAHB.

The borrowing weights support this interpretation.
When subgroup-specific shifts vary substantially, RADISH
assigns less weight than CAHB to historical information in the
$X_1=0$ subgroup, which represented approximately $69\%$ of the
HORIZON population.
With zero mean shift, the mean final borrowing weights are 
$0.173$ for RADISH and $0.416$ for CAHB.
With $g=2$, the corresponding weights are $0.095$ and $0.404$, respectively.
These smaller weights limit the influence of historical
outcomes whose conditional means differed from those of the
concurrent controls.

\begin{figure}[!ht]
\centering
\includegraphics[width=\textwidth]{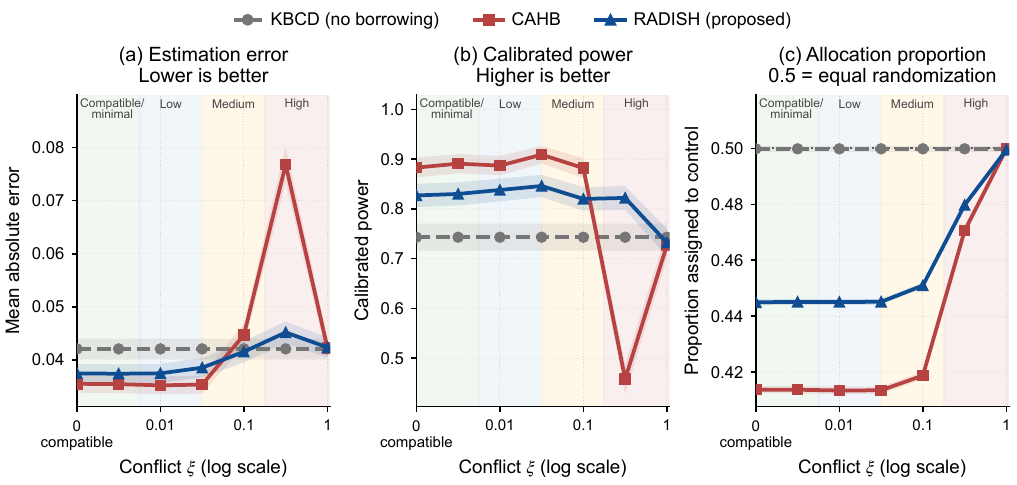}
\caption{
  Performance for HORIZON--FIT-based semi-synthetic simulation studies
  as subgroup-specific shifts increase while the mean
shift remains zero, with $N=200$, $N_H=350$, and $1000$ 
replications per scenario.
The standardized shifts are
$(b(0),b(1))/\sigma_C=4\xi\bigl\{-1,(1-p_1)/p_1\bigr\}$.
For $\xi>0$, the two shifts have opposite signs.
Positive values of $\xi$ are displayed on a logarithmic scale,
with $\xi=0$ shown at the left.
Panels show (a) mean absolute error under $\delta=0.14$,
(b) calibrated power, and (c) the proportion assigned to the control group.
Curves show point estimates, and shaded regions show pointwise $95\%$
confidence intervals.
}
\label{fig:realF2}
\end{figure}

Figure~\ref{fig:realF2} shows how borrowing and performance changed
as variation in subgroup-specific shifts increased while the mean shift remained
zero.
RADISH adapts borrowing as intended.
When historical and concurrent controls are compatible
($\xi=0$), RADISH reduces MAE from $0.042$ for KBCD to $0.037$
and increases calibrated power from $0.743$ to $0.827$.
As compatibility decreases, RADISH reduces borrowing, and its
gains over KBCD diminish.
Although RADISH's MAE is slightly higher than KBCD's in some scenarios,
RADISH retains gains in calibrated power over the range of $\xi$.
At $\xi=1$, its mean final borrowing weight is only $0.0016$,
and its estimation accuracy and calibrated power approach
those of KBCD.
The control allocation proportion is $0.499$, indicating
nearly equal allocation.
Supplementary Figure~S5 shows how borrowing and control allocation adapt to subgroup-specific
compatibility. Repeating the second set of scenarios with $N_H=30$ yielded
smaller efficiency gains, while Type~I error remained below $0.05$
across scenarios (Supplementary Figure~S6).

CAHB borrows more aggressively than RADISH.
Under compatibility, CAHB produces larger gains, with MAE
$0.035$ and calibrated power $0.883$.
However, CAHB continues to borrow in some scenarios where the
historical and concurrent conditional control means differed
substantially, resulting in large bias and low calibrated power.
At $\xi=0.316228$, its MAE is $0.077$ and Type~I error is $0.317$,
compared with $0.045$ and $0.048$ for RADISH.
At $\xi=1$, CAHB also reduces borrowing and approaches KBCD.

\section{Discussion}
\label{sec:discuss}

RADISH incorporates historical controls into covariate-adaptive randomization
and final inference, adjusting borrowing to local compatibility and historical
precision. Precise, compatible information can improve efficiency and reduce
control enrollment, while increasing uncertainty or conflict reduces borrowing.
Under the stated regularity conditions, the Studentized RADISH estimator is
asymptotically normal, supporting Wald inference. Simulations show estimation
and power gains with precise, compatible historical information and a return
toward concurrent-only performance as incompatibility increases.
Semi-synthetic studies based on HORIZON \citep{black2007once} and
FIT \citep{black1996randomised} yield qualitatively similar results.
The 24-month BMD endpoint illustrates a practical constraint: outcome-dependent allocation requires sufficiently slow recruitment or delayed updates; the simulations assume outcomes are available before the next assignment.

The local Gaussian working model currently accommodates continuous outcomes.
Extending its likelihood and conflict measure to binary, count, and
time-to-event endpoints remains to be studied.
RADISH also uses a single historical source, although multiple studies with
differing compatibility may be available
\citep{JinYin2021UIP,Lesaffre2024review,zhang2023unit}.
Source-specific compatibility assessments and weights could accommodate such
studies. The method assumes consistently defined and measured covariates
across sources and adequate historical support over the concurrent covariate
space. Extensions addressing limited overlap, inconsistent definitions, and
sparse historical information would broaden its applicability.

\backmatter
\section*{Conflict of Interest}
The authors declare no conflicts of interest.

\section*{Data Availability}
The HORIZON and FIT participant-level data are subject to the original data-use
agreements and are not distributed with this article. Access requires
authorization under the applicable data-use agreements. Simulation and analysis
code is available at \url{https://github.com/BobZhangHT/HistInfoBorrow}.

\bibliographystyle{biom}
\bibliography{references}

\section*{Supplementary Material}
Supplementary Sections~S1--S6, Table~S1, and Figures~S1--S6 referenced in
Sections~2--4 are available with this paper at the \textit{Biometrics} website
on Oxford Academic.

\end{document}


\maketitle

\setcounter{table}{0}
\setcounter{figure}{0}
\setcounter{equation}{0}
\setcounter{section}{0}
\setcounter{subsection}{0}
\setcounter{prop}{0}
\setcounter{lemma}{0}
\setcounter{theorem}{0}
\setcounter{corollary}{0}
\renewcommand{\thetable}{S\arabic{table}}
\renewcommand{\thefigure}{S\arabic{figure}}
\renewcommand{\theequation}{S\arabic{equation}}
\renewcommand{\thealgocf}{S\arabic{algocf}}
\renewcommand{\thesection}{S\arabic{section}}
\renewcommand{\thesubsection}{\thesection.\arabic{subsection}}
\renewcommand{\theprop}{S\arabic{prop}}
\renewcommand{\thelemma}{S\arabic{lemma}}
\renewcommand{\thetheorem}{S\arabic{theorem}}
\renewcommand{\thecorollary}{S\arabic{corollary}}
\renewcommand{\theHsection}{supp.\arabic{section}}
\renewcommand{\theHsubsection}{\theHsection.\arabic{subsection}}
\makeatletter
\renewcommand{\@seccntformat}[1]{%
  \expandafter\ifx\csname #1\endcsname\section
    Section~\csname the#1\endcsname\quad
  \else
    \expandafter\ifx\csname #1\endcsname\subsection
      Section~\csname the#1\endcsname\quad
    \else
      \csname the#1\endcsname\quad
    \fi
  \fi
}
\makeatother

\section{Overview}

\begin{figure}[!ht]
\centering
\resizebox{\textwidth}{!}{%
\begin{tikzpicture}[
  font=\small,
  >={Stealth[length=2.6mm]},
  stg/.style ={rectangle, rounded corners, draw, thick, align=center,
               inner sep=4pt, minimum height=18mm, text width=24mm, fill=blue!6},
  data/.style={rectangle, draw, align=center, inner sep=3pt,
               text width=17mm, fill=black!8},
  cyc/.style={rectangle, rounded corners, draw, align=center, inner sep=3pt,
               text width=21mm, minimum height=14mm, fill=white},
  arr/.style ={->, thick},
  loop/.style={->, line width=1pt, red!65!black}
]
\node[data] (hist) {Historical controls $\mathcal{D}_H$};
\node[stg, right=6mm of hist] (s1)
  {\textbf{Stage~I}\\[1pt]{\footnotesize Local historical reference}\\
   {\footnotesize $\theta(\bm{x})$ and $\hat{\tau}_H^2(\bm{x})$}\\[1pt]
   {\footnotesize fixed before enrollment}};

\node[cyc, right=10mm of s1, text width=24mm] (burn)
  {\textbf{Burn-in}\\[1pt]
   $i=1,\ldots,n_{\mathrm{init}}$\\
   randomize $1{:}1$\\ observe $Y_i$};
\node[cyc, right=6mm of burn] (a) {Enroll participant $n+1$ \\ with covariate $\bm{x}_{n+1}$};
\node[cyc, right=5mm of a]   (b) {Calculate conflict $D_n(\bm{x})$,\\ and ratio $R_n^A(\bm{x})$};
\node[cyc, right=5mm of b]   (c) {Randomize with $\phi_n(\bm{x})$, observe $Y_{n+1}$};
\draw[arr] (burn) -- (a);
\draw[arr] (a) -- (b);
\draw[arr] (b) -- (c);
\node[below=6mm of b, font=\footnotesize, align=center,
      text width=67mm, text=red!65!black] (looplbl)
  {append the new record to $\mathcal{D}_n$;\\ $n \leftarrow n+1$; repeat until $n=N$};
\draw[loop] (c.south) |- ([yshift=-3mm]a.south) -- (a.south);
\begin{scope}[on background layer]
  \node[draw, dashed, rounded corners, thick, fill=blue!3,
        inner xsep=4mm, inner ysep=4mm, fit=(burn)(a)(b)(c)(looplbl)] (s2box) {};
\end{scope}
\node[anchor=south, font=\bfseries\small] at (s2box.north)
  {Stage~II: sequential information-ratio randomization (one participant per step)};

\node[stg, right=10mm of s2box, text width=27mm] (s3)
  {\textbf{Stage~III}\\[1pt]{\footnotesize Estimate $\widehat{\Delta}_N$,}\\
   {\footnotesize linear-smoother variance, and}\\
   {\footnotesize Wald CI at $n=N$}};

\draw[arr] (hist) -- (s1);
\draw[arr] (s1.east) -- (burn.west);
\draw[arr, line width=1pt] (s2box.east) -- (s3.west)
  node[midway, above, font=\footnotesize]{$n=N$};
\end{tikzpicture}}
\caption{The RADISH trial-design procedure. Stage~I constructs the local
historical reference before enrollment. Stage~II begins with
equal-probability burn-in randomization for the first $n_{\mathrm{init}}$
participants and then updates the allocation sequentially using accrued
outcomes. Stage~III uses the completed trial data for final estimation and
Wald inference.}
\label{fig:pipeline}
\end{figure}
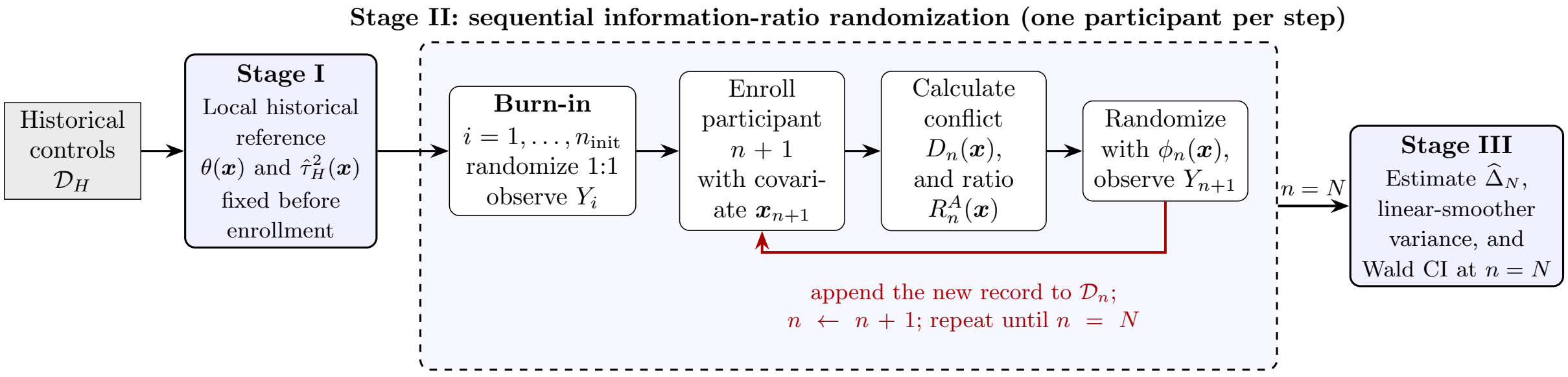

Figure~\ref{fig:pipeline} summarizes the RADISH trial-design procedure.
These supplementary materials provide additional numerical results,
followed by technical details and proofs.

In Section~\ref{supp:simulation_study}, we report the results of the simulation study.
Table~\ref{tab:metrics_null} reports
performance under the null hypothesis $\Delta=0$, and
Figure~\ref{fig:allocation} shows the allocation proportion to the control group.
Figure~\ref{fig:global_local} examines borrowing under different overall 
mean shifts and variation in subgroup-specific shifts.
Figure~\ref{fig:hist_size_sensitivity} assesses sensitivity to
historical sample size.

In Section~\ref{supp:real_sensitivity}, we report the results of the HORIZON--FIT-based semi-synthetic study.
Figure~\ref{fig:real_stratum_diagnostics} shows subgroup-specific
borrowing weights and the allocation proportion to the control group, and
Figure~\ref{fig:real_sensitivity_efficiency} examines sensitivity
to historical sample size.

In Section~\ref{supp:stage3_details}, we provide additional details for Stage~III.
In Section~\ref{supp:asymptotic_theory}, we provide the asymptotic theory and proofs.
Section~\ref{supp:algorithms} gives detailed Algorithms~\ref{alg:stage1}--\ref{alg:stage3} for the three-stage overview in Algorithm~1 of the main text.

\section{Additional Results for the Simulation Study}
\label{supp:simulation_study}

\subsection{Null-Hypothesis Results and Control Allocation}

\begin{figure}[!ht]
\centering
\includegraphics[width=\textwidth]{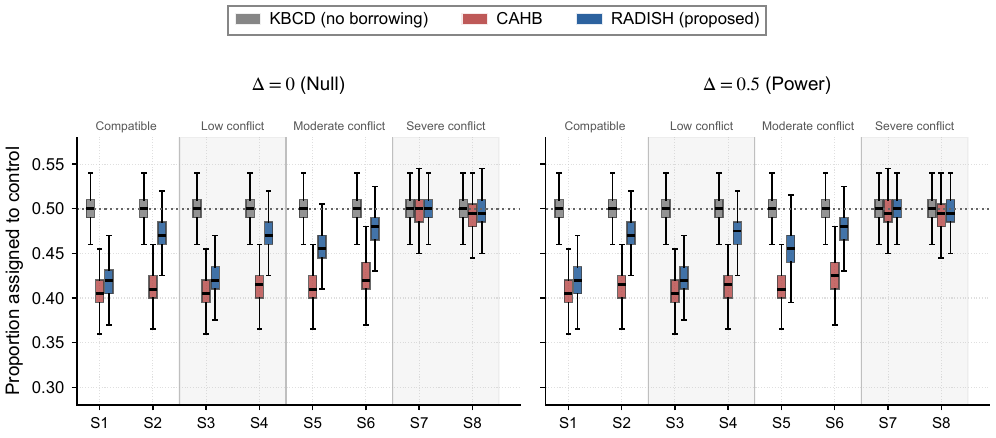}
\caption{Overall proportion assigned to the control group across $M=1000$ paired
replications under $\Delta=0$ and $\Delta=0.5$, with $N_H=N=200$. The eight scenarios cross
$(b(0),b(1))=(0,0),(0.04,0.12),(0.25,0.75),(1.5,2.5)$ with
$\sigma_H=0.5,3.0$, respectively; colors identify KBCD, CAHB, and RADISH.
Boxes show the interquartile range with the median line, and whiskers use the
standard 1.5-IQR rule. The horizontal reference at 0.5 denotes equal
randomization; values below 0.5 indicate allocating fewer participants to the control group,
whose interpretation depends on historical--concurrent compatibility. KBCD is the concurrent-only reference, CAHB
is the borrowing comparator, and RADISH is the proposed locally adaptive method.}
\label{fig:allocation}
\end{figure}

In the main text, we reported the results of the simulation study under the alternative hypothesis $\Delta=0.5$.
Table~\ref{tab:metrics_null} summarizes estimation accuracy,
confidence-interval coverage and width, and borrowing under
the null hypothesis $\Delta=0$. 
Figure~\ref{fig:allocation} compares the proportions assigned to the control group
across the eight simulation scenarios.

\begin{table}[!ht]
\centering
\caption{
  Estimation, interval, allocation, and borrowing performance under
$\Delta=0$, based on $M=1000$ paired replications with $N_H=N=200$.
Scenarios S1--S8 combine standardized historical shifts
$(b(0),b(1))=(0,0),(0.04,0.12),(0.25,0.75),(1.5,2.5)$ with
$\sigma_H=0.5,3.0$, respectively. RMSE is root mean squared error; Width is
mean 95\% interval width; Alloc. Proportion is the proportion assigned to the control group;
$\overline W_N^F$ is the mean Stage~III borrowing weight.}
\label{tab:metrics_null}
\sisetup{round-mode=places, round-precision=3, table-format=-1.3}
\resizebox{0.92\textwidth}{!}{%
\begin{tabular}{l l S[table-format=-1.3] S[table-format=1.4, round-precision=4] S[table-format=1.3] S[table-format=1.4, round-precision=4] S[table-format=1.3] S[table-format=1.3]}
\toprule
{Scenario} & {Method} & {Bias} & {RMSE} & {Coverage} & {Width} & {Alloc.\ Proportion} & {$\overline W_N^F$} \\
\midrule
\multirow{3}{*}{S1: Compatible, $\sigma_H\!=\!0.5$}
  & KBCD   & -0.003 & 0.1634 & 0.945 & 0.6174 & 0.500 & 0.000 \\
  & CAHB   & -0.004 & 0.1281 & 0.948 & 0.4902 & 0.407 & 0.512 \\
  & RADISH & -0.003 & 0.1424 & 0.969 & 0.6188 & 0.419 & 0.383 \\
\midrule
\multirow{3}{*}{S2: Compatible, $\sigma_H\!=\!3.0$}
  & KBCD   & -0.003 & 0.1634 & 0.945 & 0.6174 & 0.500 & 0.000 \\
  & CAHB   & -0.004 & 0.1676 & 0.954 & 0.6536 & 0.413 & 0.478 \\
  & RADISH & -0.000 & 0.1571 & 0.952 & 0.6159 & 0.473 & 0.127 \\
\midrule
\multirow{3}{*}{S3: Small, $\sigma_H\!=\!0.5$}
  & KBCD   & -0.003 & 0.1634 & 0.945 & 0.6174 & 0.500 & 0.000 \\
  & CAHB   & -0.044 & 0.1354 & 0.924 & 0.4902 & 0.407 & 0.512 \\
  & RADISH & -0.019 & 0.1448 & 0.962 & 0.6186 & 0.421 & 0.369 \\
\midrule
\multirow{3}{*}{S4: Small, $\sigma_H\!=\!3.0$}
  & KBCD   & -0.003 & 0.1634 & 0.945 & 0.6174 & 0.500 & 0.000 \\
  & CAHB   & -0.037 & 0.1714 & 0.949 & 0.6535 & 0.414 & 0.477 \\
  & RADISH & -0.007 & 0.1559 & 0.952 & 0.6160 & 0.473 & 0.126 \\
\midrule
\multirow{3}{*}{S5: Moderate, $\sigma_H\!=\!0.5$}
  & KBCD   & -0.003 & 0.1634 & 0.945 & 0.6174 & 0.500 & 0.000 \\
  & CAHB   & -0.232 & 0.2774 & 0.513 & 0.4972 & 0.412 & 0.486 \\
  & RADISH & -0.028 & 0.1645 & 0.938 & 0.6171 & 0.457 & 0.135 \\
\midrule
\multirow{3}{*}{S6: Moderate, $\sigma_H\!=\!3.0$}
  & KBCD   & -0.003 & 0.1634 & 0.945 & 0.6174 & 0.500 & 0.000 \\
  & CAHB   & -0.161 & 0.2429 & 0.817 & 0.6459 & 0.424 & 0.417 \\
  & RADISH & -0.018 & 0.1607 & 0.943 & 0.6161 & 0.479 & 0.087 \\
\midrule
\multirow{3}{*}{S7: Severe, $\sigma_H\!=\!0.5$}
  & KBCD   & -0.003 & 0.1634 & 0.945 & 0.6174 & 0.500 & 0.000 \\
  & CAHB   & -0.007 & 0.1717 & 0.929 & 0.6164 & 0.497 & 0.004 \\
  & RADISH & -0.004 & 0.1636 & 0.940 & 0.6174 & 0.500 & 0.001 \\
\midrule
\multirow{3}{*}{S8: Severe, $\sigma_H\!=\!3.0$}
  & KBCD   & -0.003 & 0.1634 & 0.945 & 0.6174 & 0.500 & 0.000 \\
  & CAHB   & -0.026 & 0.1789 & 0.921 & 0.6176 & 0.493 & 0.028 \\
  & RADISH & -0.005 & 0.1629 & 0.943 & 0.6173 & 0.498 & 0.004 \\
\bottomrule
\end{tabular}}
\end{table}

\subsection{Mean Shift and Variation in Subgroup-Specific Shifts}
\label{supp:global_local}

To distinguish the marginal shift from local heterogeneity, we write
$b(X_1)=g+\ell s(X_1)$ with $P(X_1=1)=1/3$, $s(0)=-1/2$, and $s(1)=1$, so
$E\{s(X_1)\}=0$ and $E\{b(X_1)\}=g$. Thus $g$ is the marginal historical
shift and $\ell$ controls how strongly that shift varies between the two
covariate strata. Each ordered pair below is $(b(0),b(1))$: its first entry
is the historical mean shift for $X_1=0$, and its second entry is the shift for
$X_1=1$, both relative to the concurrent control mean. For example,
$(-0.05,0.10)$ means a shift of $-0.05$ in the $X_1=0$ stratum and $0.10$ in
the $X_1=1$ stratum; these shifts average to zero under the target distribution.

Crossing $g\in\{0,0.6\}$ with $\ell\in\{0.1,1.2\}$ gives four regimes:
weak marginal shift and weak local heterogeneity $(-0.05,0.10)$, strong
marginal shift and weak local heterogeneity $(0.55,0.70)$, weak marginal shift
and strong local heterogeneity $(-0.60,1.20)$, and strong marginal shift and
strong local heterogeneity $(0,1.80)$. In the last regime, the overall shift
and the between-stratum contrast are both strong: $X_1=0$ remains exactly
compatible, whereas $X_1=1$ has a shift of $1.80$. Every regime remains covariate
dependent, while $g$ and $\ell$ vary independently. Here $N_H=N=200$ and
$\sigma_H=0.5$; the remaining settings match the primary simulation, and each
cell uses 1000 paired replications.

\begin{figure}[!th]
\centering
\includegraphics[width=\textwidth]{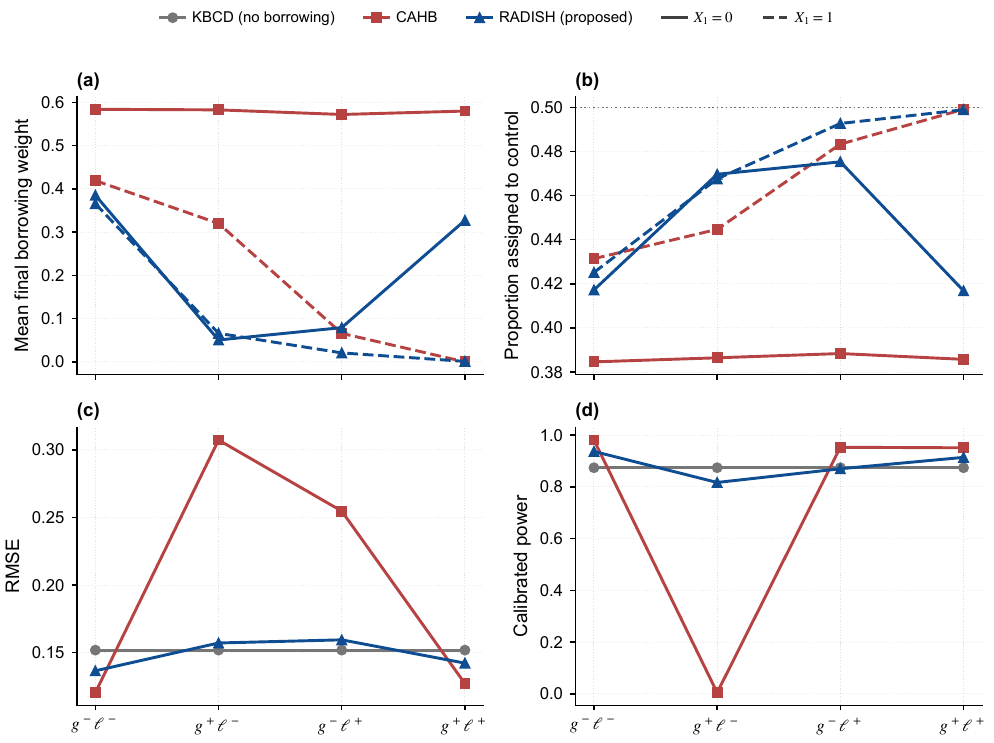}
\caption{Marginal-shift-by-local-heterogeneity diagnostic based on $1000$
paired replications with $N_H=N=200$ and $\sigma_H=0.5$. The four
horizontal-axis categories cross weak/strong marginal
shift with weak/strong local heterogeneity. The horizontal-axis labels use $g$
for the marginal shift $g=E\{b(X_1)\}$ and $\ell$ for the heterogeneity
parameter. In plot order, the four pairs $(b(0),b(1))$ are
$(-0.05,0.10)$, $(0.55,0.70)$, $(-0.60,1.20)$, and $(0,1.80)$.
Here $\Delta=0.5$. In the $g^+\ell^+$ category,
$(b(0),b(1))=(0,1.80)$, so $X_1=0$ remains compatible even though the
marginal shift and between-stratum heterogeneity are strong. Panels show (a) terminal borrowing weights by
$X_1$ stratum, (b) the allocation proportion to the control group by stratum, (c) RMSE under
$\Delta=0.5$, and (d) calibrated power using the cell-specific empirical null
critical value. Solid and dashed curves denote $X_1=0$ and $X_1=1$.
}
\label{fig:global_local}
\end{figure}

Figure~\ref{fig:global_local} illustrates the behavior of the combined
score: broadly distributed conflict activates the marginal safeguard, while
the local score distinguishes covariate strata when marginal averaging masks
heterogeneous conflict. In the two regimes where only one conflict axis is
strong, RADISH remains close to the concurrent-only RMSE and calibrated-power
references, whereas CAHB is sensitive to the unresolved conflict. In the
strong-marginal-shift, strong-local-heterogeneity regime, RADISH retains
mean weights $0.328/0.001$ for $X_1=0/1$. The nonzero weight in the first
stratum is expected because $b(0)=0$, whereas the nearly zero second weight
reflects the severe local conflict $b(1)=1.80$. In the weak-marginal-shift,
strong-local-heterogeneity regime, the corresponding weights are
$0.080/0.021$. These results support the rationale
for combining local and marginal conflict through a local gate.

\subsection{Sensitivity Analysis of the Historical Sample Size}
\label{supp:hist_size_sensitivity}

\begin{figure}[!ht]
\centering
\includegraphics[width=\textwidth]{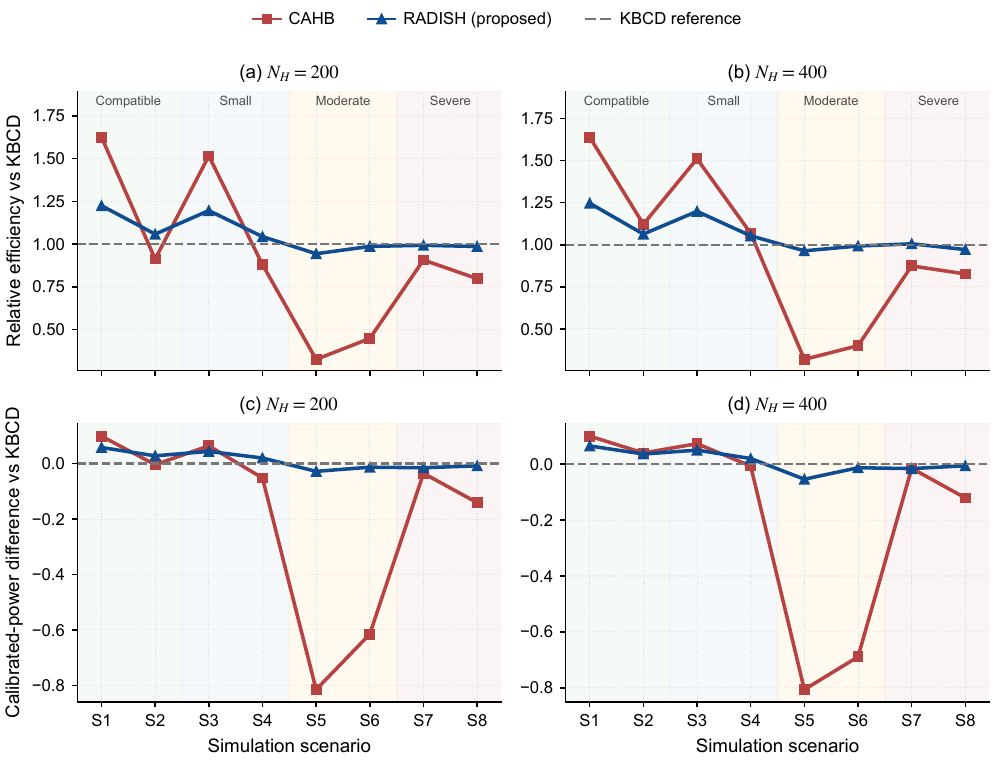}
\caption{Sensitivity of the simulation results to doubling the historical
sample size from $N_H=200$ to $400$, based on $1000$ paired replications per
cell with $N=200$. Columns correspond to the two historical sample sizes.
Panels (a) and (b) show relative efficiency,
$(\mathrm{RMSE}_{\mathrm{KBCD}}/\mathrm{RMSE})^2$, so values above one favor
the borrowing method. Panels (c) and (d) show calibrated-power differences
relative to KBCD, so values above zero favor the borrowing method. The dashed
lines are the KBCD references. S1--S2 are compatible, S3--S4 have small local
conflict, S5--S6 have moderate local conflict, and S7--S8 have severe local
conflict; within each pair, the first and second scenarios use
$\sigma_H=0.5$ and $3.0$, respectively. All nonzero conflicts depend on
$X_1$. KBCD is concurrent-only, CAHB is the borrowing comparator, and RADISH
is the proposed method.}
\label{fig:hist_size_sensitivity}
\end{figure}

The primary simulation fixes $N_H=200$. We repeated the eight scenarios
with $N_H=400$ using nested historical cohorts and paired concurrent-trial
random-number streams, as shown in Figure~\ref{fig:hist_size_sensitivity}.
Doubling the historical cohort produces a modest
additional efficiency gain under compatible and small-conflict settings, whereas
RADISH remains close to KBCD under moderate and severe conflict because its
borrowing weights decrease as incompatibility becomes detectable. Thus the
larger historical cohort increases the gains from compatible borrowing while
preserving the return toward concurrent-only analysis as conflict increases.

\section{Additional Results for the Semi-Synthetic Simulation Study}
\label{supp:real_sensitivity}

\begin{figure}[!ht]
\centering
\includegraphics[width=\textwidth]{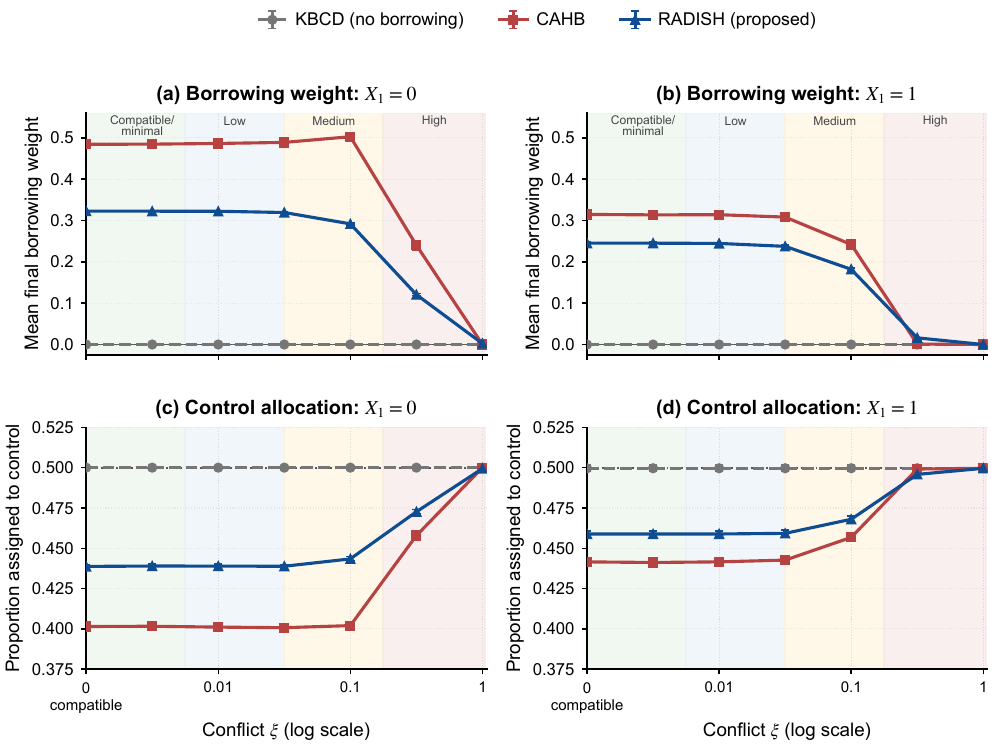}
\caption{Subgroup-specific borrowing diagnostics for the HORIZON--FIT-based
semi-synthetic simulation, with $N=200$, $N_H=350$, $\delta=0.14$,
and $1000$ replications per scenario.
The standardized shifts are
$(b(0),b(1))/\sigma_C=4\xi\bigl\{-1,(1-p_1)/p_1\bigr\}$,
where $p_1\approx0.310$.
Their magnitudes increase with $\xi$ while the mean shift remains zero.
Panels (a) and (b) show mean final borrowing weights for the
$X_1=0$ and $X_1=1$ subgroups, respectively.
KBCD provides the zero-borrowing reference.
Panels (c) and (d) show the corresponding proportions assigned
to the control group.
The dotted line marks equal allocation at $0.5$.
Error bars show pointwise normal-approximation $95\%$ confidence
intervals.
Positive values of $\xi$ are displayed on a logarithmic scale,
with $\xi=0$ shown at the left.
}
\label{fig:real_stratum_diagnostics}
\end{figure}

Figure~\ref{fig:real_stratum_diagnostics} complements the aggregate
operating characteristics in the main text by separating the two $X_1$
regions. It shows how subgroup-specific borrowing and randomization produce the
overall response along the localized-conflict path.

\begin{figure}[!ht]
\centering
\includegraphics[width=\textwidth]{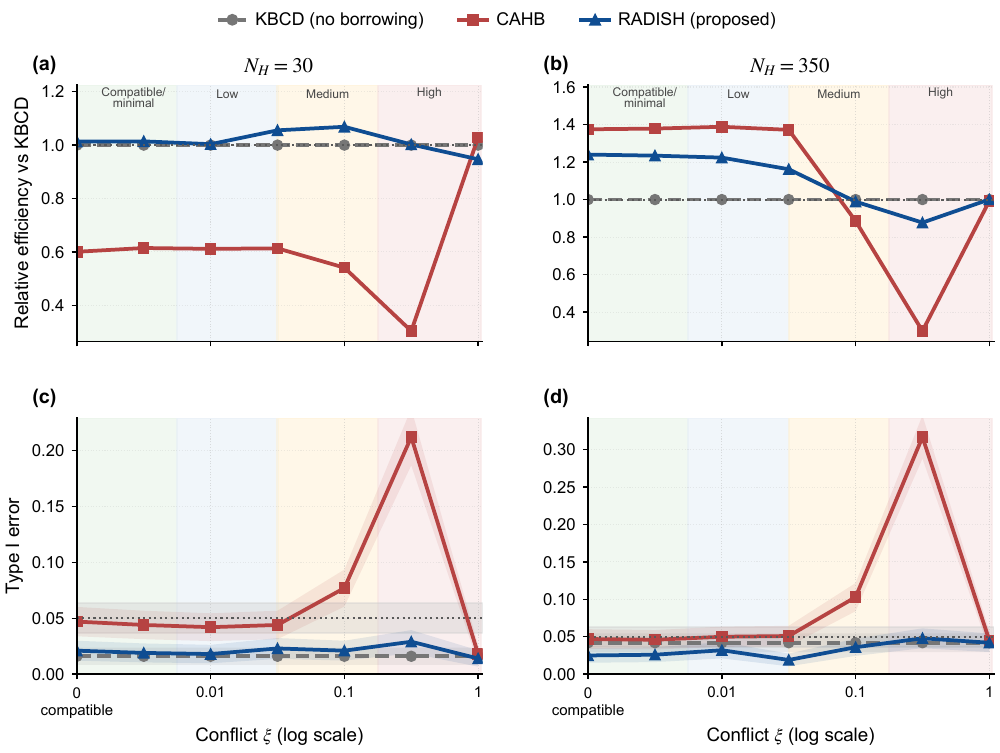}
\caption{
  Sensitivity analysis of the historical sample size $N_H$
   for the HORIZON--FIT-based semi-synthetic simulation
as subgroup-specific shifts increase while the mean shift remains
zero, with $N=200$ and $1000$ replications per scenario.
The left and right columns use $N_H=30$ and $N_H=350$, respectively.
The standardized shifts are
$(b(0),b(1))/\sigma_C=4\xi\bigl\{-1,(1-p_1)/p_1\bigr\}$,
where $p_1\approx0.310$.
For $\xi>0$, the two shifts have opposite signs.
Positive values of $\xi$ are displayed on a logarithmic scale,
with $\xi=0$ shown at the left.
Panels (a) and (b) show relative efficiency under $\delta=0.14$,
defined as $(\mathrm{RMSE}_{\mathrm{KBCD}}/\mathrm{RMSE})^2$.
Panels (c) and (d) show empirical Type~I error.
Dotted lines mark relative efficiency $1$ and Type~I error $0.05$.
The gray band shows the normal-approximation $95\%$ Monte Carlo
confidence interval under Type~I error $0.05$.}
\label{fig:real_sensitivity_efficiency}
\end{figure}

The main HORIZON--FIT conflict path uses $N_H=350$. This sensitivity
analysis repeats the same seven levels
$\xi\in\{0,0.003162,0.01,0.031623,0.1,0.316228,1\}$ at
$N_H=30$, as shown in Figure~\ref{fig:real_sensitivity_efficiency}.
Along the path,
$(b(0),b(1))/\sigma_C=4\xi\{-1,(1-p_1)/p_1\}$, so every positive $\xi$
remains covariate dependent while the marginal shift is zero. The comparison
assesses whether the local-conflict response
persists at lower historical precision.

\section{Additional Computational Details}
\label{supp:stage3_details}

\subsection{Historical Reference Variance}
Because $\theta(\bm{x})$ is a linear combination of conditionally independent
observations, its variance has a closed form under the working model after
conditioning on the historical covariates $\{\bm{X}_j^H\}_{j=1}^{N_H}$:
\begin{equation*}
\tau_H^2(\bm{x}) \;=\; \sigma_H^2(\bm{x}) \sum_{j=1}^{N_H} \{S_{H,j}(\bm{x})\}^2
\;=\; \frac{\sigma_H^2(\bm{x})}{n_H^{\mathrm{eff}}(\bm{x})},
\qquad
n_H^{\mathrm{eff}}(\bm{x}) \;=\; \frac{\bigl\{\sum_j w_j^{H}(\bm{x})\bigr\}^2}
                                       {\sum_j \{w_j^{H}(\bm{x})\}^2},
\end{equation*}
where $n_H^{\mathrm{eff}}(\bm{x})$ is the local effective historical sample size for constructing $\theta(\bm{x})$.
The unknown $\sigma_H^2(\bm{x})$ is replaced by the weighted residual estimator
\[
\hat{\sigma}_H^2(\bm{x})
=\allowbreak
[\sum_{j}w_j^{H}(\bm{x})\{Y_j^H-\theta(\bm{x})\}^2]/
\{\sum_{j}w_j^{H}(\bm{x})\}.
\]
Substituting this estimator gives
$\hat{\tau}_H^2(\bm{x})=\hat{\sigma}_H^2(\bm{x})/n_H^{\mathrm{eff}}(\bm{x})$.

\subsection{Information-Ratio Interpretation}
The increment above one reflects both the precision and compatibility of the
historical reference. The variance ratio
$
\{\hat{\sigma}_{0,n}^2(\bm{x})/
\widetilde n_{0,n}^{\mathrm{eff}}(\bm{x})\}/
{\mathrm{SE}_n^2(\bm{x})}
$
compares the uncertainty in the concurrent local control mean with
the total uncertainty in the historical--concurrent comparison.
For a fixed concurrent sampling variance, this ratio is larger when the historical
reference is estimated more precisely and decreases as its estimation
uncertainty increases. The factor $\exp\{-D_n(\bm{x})\}$ further discounts
historical information as local incompatibility increases. In particular, it
equals one within the tolerance region where $D_n(\bm{x})=0$ and decreases as
the historical and concurrent local means become less compatible.

Thus, a larger $R_n^A(\bm{x})$ corresponds to greater historical borrowing
when the historical reference is both precise and locally compatible with the
concurrent controls. As the historical reference becomes less precise or less
compatible, $R_n^A(\bm{x})$ approaches one, corresponding to progressively
weaker historical borrowing.

\subsection{Residual Variance Calculations}
We estimate these source-specific outcome variances using residuals from the
same kernel smoothing procedure used to construct the point estimator. The
ordinary within-source sample variance is not appropriate here because it
contains both variation in the conditional mean across covariates and residual
variation around that mean.

For $z\in\{0,1\}$, let $\bm X_z$ and $\bm Y_z$ denote the covariate matrix and
outcome vector for concurrent participants in arm $z$, and define the
within-arm smoothing matrix
\[
\bm S_{zz}
=
\operatorname{normRows}
\left\{
{\bm K}_{\bm h}(\bm X_z,\bm X_z)
\right\},
\]
where $\operatorname{normRows}(\bm S)$ divides each row of $\bm S$ by its row
sum.
Here, ${\bm K}_{\bm h}$ is the pairwise kernel matrix whose
$(i,j)$th entry is
$
\{\bm K_{\bm h}(\bm X_z,\bm X_z)\}_{ij}
=
K_{\bm h}(\bm x_i-\bm x_j),
$
where $\bm x_i$ and $\bm x_j$ denote the corresponding row vectors.

The fitted outcomes are $\bm S_{zz}\bm Y_z$, and we estimate the residual
variance by
\[
\widehat{\sigma}_z^2
=
\frac{
\|\bm Y_z-\bm S_{zz}\bm Y_z\|_2^2
}{
\max\left\{
N_z
-2\operatorname{tr}(\bm S_{zz})
+\operatorname{tr}(\bm S_{zz}^{\top}\bm S_{zz}),
1
\right\}
},
\qquad z\in\{0,1\}.
\]
The denominator is the residual degrees of freedom associated with the linear
smoother, since
$
\operatorname{tr}
\left[
(\bm I_{N_z}-\bm S_{zz})^\top
(\bm I_{N_z}-\bm S_{zz})
\right]
=
N_z
-2\operatorname{tr}(\bm S_{zz})
+\operatorname{tr}(\bm S_{zz}^{\top}\bm S_{zz}).
$
The lower bound of one is a finite-sample safeguard that ensures a positive
residual degrees-of-freedom denominator.

The historical residual variance is estimated analogously. Let $\bm X_H$ and
$\bm Y_H$ denote the historical covariate matrix and outcome vector, and define
$
\bm S_{HH}
=
\operatorname{normRows}
\left\{
{\bm K}_{\bm h}(\bm X_H,\bm X_H)
\right\}.
$
Then
\[
\widehat{\sigma}_H^2
=
\frac{
\|\bm Y_H-\bm S_{HH}\bm Y_H\|_2^2
}{
\max\left\{
N_H
-2\operatorname{tr}(\bm S_{HH})
+\operatorname{tr}(\bm S_{HH}^{\top}\bm S_{HH}),
1
\right\}
}.
\]
Here $\widehat{\sigma}_H^2$ is the source-wide residual variance used for final
inference and is distinct from the local Stage~I variance
$\widehat{\sigma}_H^2(\bm x)$.
For numerical evaluation, residual variances and the computed values of
$\widehat V_{\mathrm{lin}}$ and $\widehat V_C$ are floored at
$\epsilon_v=10^{-12}$ before forming their final maximum.

\subsection{Implementation Properties}
For Scott's rule, the historical covariate standard deviation
$\hat\sigma_{H,k}$ is replaced by one if it is less than $10^{-12}$,
keeping the bandwidth positive. Several properties of the kernel calculations
are also useful for implementation and interpretation. Multiplying the kernel by any positive constant leaves all
row-normalized smoothing weights and local effective sample sizes unchanged.
The same multiplier applies to both arm-specific kernel masses in the
randomization probability and therefore cancels. Thus a common rescaling of the
kernel does not change the local means, information ratios, borrowing weights,
randomization probabilities, or final estimator.

Within each data source, the outcome vector, covariate matrix, and smoothing
coefficients use the same observation ordering. This convention ensures that
the entries of the cumulative coefficient vectors multiply the corresponding
outcomes in the weighted-sum representation of the estimator.

The trace in the residual degrees of freedom has a direct working-model
interpretation. For a fixed smoother $\bm S$ and
$\bm Y=\bm\mu+\bm\varepsilon$, with independent mean-zero homoskedastic errors of variance
$\sigma^2$,
\[
\E\{\|(\bm I-\bm S)\bm Y\|_2^2\}
=
\|(\bm I-\bm S)\bm\mu\|_2^2
+
\sigma^2\operatorname{tr}\{(\bm I-\bm S)^\top(\bm I-\bm S)\}.
\]
The trace therefore normalizes the noise contribution to the residual sum of
squares; consistency also requires the squared mean-approximation error,
divided by the residual degrees of freedom, to vanish. For
RADISH, conditions (C4) and (C6) in Section~\ref{supp:asymptotic_framework} transfer this argument from the
outcome-independent oracle smoother to the realized adaptive smoother.

The probability floor bounds the conflict scores whenever their standardized
discrepancies are defined, and the local variance floors keep the local standard
errors positive. Well-definedness of the marginal discrepancy is imposed
separately in condition (C4). The resulting exponential discount is strictly
positive, although it can be numerically negligible under severe conflict.


\section{Asymptotic Theory and Proofs}
\label{supp:asymptotic_theory}
\subsection{Large-Sample Properties}
\label{supp:large_sample_properties}

The asymptotic analysis conditions on both the historical dataset and the
enrolled covariates. Under this conditional formulation, $\Delta_N$ in
Equation~(1) of the main text is fixed, and repeated-trial uncertainty arises
from sequential randomization and the outcome errors. The working variance
therefore quantifies uncertainty for the treatment effect in the enrolled
cohort conditional on the observed covariate distribution.

The central limit theorem accounts for the dependence of RADISH allocation
probabilities and terminal borrowing weights on concurrent control outcomes.
The smoothing coefficients in $\widehat\Delta_N$ are therefore dependent on the
outcomes they multiply. We separate this dependence from the outcome variation
used for inference through a hypothetical concurrent-only version of RADISH.
This oracle retains the covariate-adaptive sequential structure and sets
historical borrowing to zero in allocation and final estimation. Comparing the
actual and oracle paths under common randomization draws isolates the effect of
historical borrowing on allocation and estimation. Under the coupling rates
stated below, the oracle large-sample distribution also describes the RADISH
estimator to first order.

Let $U_1,\ldots,U_N$ be independent standard uniform variables, independent of
the historical data, enrolled covariates, and potential-outcome errors. These
variables represent the random draws used to implement the sequential
allocation probabilities, so the actual assignment has the exact
representation $Z_i=\mathbf 1\{U_i\leq\phi_{i-1}(\bm X_i)\}$. Using the same
$U_i$ for the actual and oracle paths couples their assignments participant by
participant. Conditional on their allocation probabilities, the two assignments
differ only when $U_i$ falls between those probabilities.

On this common probability space, the oracle uses the same burn-in and the same
$U_i$, but sets the historical information ratio to one. After burn-in, its
treatment probability is
\[
\phi_{i-1}^{\circ}(\bm X_i)
=
\frac{
\left\{
\sum_{k<i}K_{\bm h}(\bm X_k-\bm X_i)
\mathbf 1(Z_k^{\circ}=0)
\right\}^{2}
}{
\left\{
\sum_{k<i}K_{\bm h}(\bm X_k-\bm X_i)
\mathbf 1(Z_k^{\circ}=0)
\right\}^{2}
+
\left\{
\sum_{k<i}K_{\bm h}(\bm X_k-\bm X_i)
\mathbf 1(Z_k^{\circ}=1)
\right\}^{2}
},
\]
and $Z_i^{\circ}=\mathbf 1\{U_i\leq
\phi_{i-1}^{\circ}(\bm X_i)\}$. The oracle probability depends only on the
covariates and preceding oracle assignments.

Applying the concurrent part of Stage~III to the oracle path gives
$\bm\alpha_{1N}^{\circ},\bm\alpha_{0N}^{\circ}\in\mathbb R^N$. Each vector has
one position for every enrolled participant. The $i$th entry of
$\bm\alpha_{zN}^{\circ}$ is zero if oracle participant $i$ is not assigned to arm
$z$ and otherwise equals that participant's averaged kernel coefficient. Using the
common participant index $1,\ldots,N$, the oracle estimator is
$\widehat\Delta_N^{\circ}=\sum_{i=1}^N\alpha_{1N,i}^{\circ}Y_i(1)
-\sum_{i=1}^N\alpha_{0N,i}^{\circ}Y_i(0)$.
Conditional on the enrolled covariates and the variables $U_1,\ldots,U_N$, the
oracle coefficients are fixed with respect to the outcome errors. The stochastic
part of $\widehat\Delta_N^{\circ}$ is therefore a weighted sum of independent,
mean-zero errors, which permits a conditional weighted central limit theorem.
The proof then establishes first-order equivalence between the RADISH estimator
and this oracle estimator.

\begin{prop}[Large-sample reversion to concurrent-only design and analysis]
\label{prop:radish_design_reversion}\leavevmode\par\nobreak
Suppose conditions \condref{C1}, \condref{C2}, and \condref{C4} in
Section~\ref{supp:asymptotic_framework} hold. Then
\[
\frac1N\sum_{i=1}^N\mathbf 1(Z_i\ne Z_i^{\circ})
\xrightarrow{p}0,
\qquad
\frac1N\sum_{i=1}^NW_N^F(\bm X_i)
\xrightarrow{p}0,
\]
and
\[
\widehat\Delta_N-\widehat\Delta_N^{\circ}=o_p(N^{-1/2}).
\]
\end{prop}

Proposition~\ref{prop:radish_design_reversion} has implications for both trial
design and inference. Because the two paths use the same randomization
variables, the first limit measures the effect of historical borrowing on
treatment allocation. This limit shows that borrowing changes
the assignments of a vanishing proportion of participants. The second
limit shows that the average contribution of the fixed historical dataset to
the final control analysis vanishes as concurrent information accumulates. The
third limit gives the rate needed for inference: the difference between the
RADISH and oracle estimators is smaller than the root-$N$ sampling scale,
allowing them to share the same first-order limit under the variance conditions
stated below.
Together, these limits describe how borrowing changes with concurrent
enrollment. Historical information can influence allocation and contribute
additional precision at trial sizes used in practice. With a fixed historical
dataset, its contribution vanishes as concurrent information accumulates. The
proposition links this gradual return to concurrent-only analysis with the
proof of asymptotic normality. The empirical
results quantify the finite-sample contribution of borrowing.

Theorem~1 of the main text combines the conditional central
limit theorem for the oracle with first-order equivalence of the RADISH and
oracle estimators, stability of their coefficient vectors, and consistency of
the working variance. Thus the proof explicitly tracks the outcome dependence
of the RADISH allocation, borrowing weights, and realized coefficients. The
regularity conditions and proofs are given in Sections~\ref{supp:asymptotic_framework}--\ref{supp:studentized_proof}.

\begin{corollary}[Large-sample validity of Wald inference]
\label{cor:radish_wald_validity}\leavevmode\par\nobreak
Under the conditions of Theorem~1, the interval in Equation~(8) of the main
text has conditional limiting coverage $1-\alpha$.
The two-sided Wald test has conditional limiting size $\alpha$ under
$H_0:\Delta_N=0$.
\end{corollary}

\subsection{Asymptotic Framework and Regularity Conditions}
\label{supp:asymptotic_framework}

The theory considers a triangular sequence of concurrent trials indexed by
$N$, conditional on the historical dataset $\mathcal D_H$ observed before
enrollment and on the enrolled covariates $\bm X_{1:N}$. Thus $N_H$, the
historical outcomes, and $\Delta_N$ are fixed in the conditional analysis, and
the variance targets randomization and outcome uncertainty for the enrolled
cohort. The argument also conditions on the auxiliary randomization variables
when fixing the oracle coefficients.
Unless stated otherwise, all probabilities, expectations, stochastic
orders, and limits in this asymptotic analysis are understood under the law
conditional on $(\mathcal D_H,\bm X_{1:N})$.
Let $\mathcal F_{i-1}^{X}=\sigma\{\mathcal D_H,
(\bm X_k,Z_k,Y_k)_{k<i},\bm X_i\}$ denote the information available
immediately before participant $i$ is randomized, augmented by the incoming
covariate $\bm X_i$.
The actual RADISH allocation is nonanticipating because $Z_i$ is generated
before $Y_i$ is observed.  

For the proofs, embed the two concurrent coefficient vectors in
$\mathbb R^N$. Define
\[
\widehat\alpha_{zN,i}
=
\begin{cases}
(\bm a_z)_{\sum_{k=1}^i\mathbf 1(Z_k=z)}, & Z_i=z,\\
0, & Z_i\ne z,
\end{cases}
\qquad z\in\{0,1\},
\]
and set $\widehat{\bm\alpha}_{HN}=\bm a_H\in\mathbb R^{N_H}$. The zero entries
place both concurrent coefficient vectors on the common participant index, while
the estimator continues to use the observed outcome from each assigned arm.
With this
notation,
\begin{equation*}
\widehat\Delta_N
=
\sum_{i=1}^N\widehat\alpha_{1N,i}Y_i(1)
-
\sum_{i=1}^N\widehat\alpha_{0N,i}Y_i(0)
-
\sum_{j=1}^{N_H}\widehat\alpha_{HN,j}Y_j^H.
\end{equation*}
Exactly one of the first two terms is observed for each concurrent participant, because
the unused coefficient is zero.

Conditions \condref{C1}--\condref{C6} are organized in the order in which they
enter the proof. Condition \condref{C4} is stated in terms of
allocation-probability perturbations and normalized kernel rows. From these conditions, Lemma
\ref{lem:supp_oracle_equivalence} establishes the first-order oracle equivalence
used by the main theorem.

\phantomsection\label{cond:C1}
\paragraph{(C1) Concurrent outcomes, sequential randomization, and moments}
Concurrent potential outcomes satisfy
$Y_i(z)=\mu_z(\bm X_i)+\varepsilon_i(z)$ for $z\in\{0,1\}$, and the observed
concurrent outcome obeys consistency,
$Y_i=Z_iY_i(1)+(1-Z_i)Y_i(0)$.
Conditional on $(\mathcal D_H,\bm X_{1:N})$, the concurrent error pairs are
independent across participants and satisfy
$\E\{\varepsilon_i(z)\mid\mathcal D_H,\bm X_{1:N}\}=0$. Dependence between
$\varepsilon_i(0)$ and $\varepsilon_i(1)$ for the same concurrent participant is
unrestricted. For some $q>2$ and a fixed finite constant $C$, the potential
outcomes satisfy, almost surely, the uniform conditional moment bound
$\sup_{i,z}\E\{|Y_i(z)|^q\mid\mathcal D_H,\bm X_{1:N}\}\leq C$.
The inequality $|a-b|^q\leq2^{q-1}(|a|^q+|b|^q)$, together with
Jensen's inequality, gives the corresponding bound for $\varepsilon_i(z)$.
To match the working variance in Equation~(7) of the main text, assume
source-wise homoskedasticity:
$\Var\{\varepsilon_i(z)\mid\mathcal D_H,\bm X_{1:N}\}=\sigma_z^2$ for
$z\in\{0,1\}$,
where both variances are bounded away from zero and infinity.  Treatment is
assigned nonanticipatively: $\Pr(Z_i=1\mid\mathcal F_{i-1}^{X})
=\phi_{i-1}(\bm X_i)$ and
$Z_i\indep\{Y_i(0),Y_i(1)\}\mid\mathcal F_{i-1}^{X}$.
The common randomized burn-in supplies the initial concurrent observations;
condition (C1) requires positive support for both arms before the adaptive rule
is used.

\phantomsection\label{cond:C2}
\paragraph{(C2) Outcome-independent concurrent-only KBCD oracle}
Let $\mathcal H_N$ denote the sigma-field generated by $\mathcal D_H$ and
$\bm X_1,\ldots,\bm X_N$.
The common bandwidth and every allocation input fixed before the first
randomization are \(\mathcal H_N\)-measurable.
Let $U_1,\ldots,U_N$ be independent standard uniform variables that are jointly
independent of $\mathcal D_H$, $\bm X_{1:N}$, and all potential-outcome errors.
The actual assignment can be represented exactly as
$Z_i=\mathbf 1\{U_i\leq\phi_{i-1}(\bm X_i)\}$.  Starting from the same burn-in,
define the oracle design by setting the historical information ratio to one in
the Stage~II allocation rule and using the same $U_i$.  Its assignment is
$Z_i^{\circ}=\mathbf 1\{U_i\leq\phi_{i-1}^{\circ}(\bm X_i)\}$.  After burn-in,
the oracle probability is
\[
\phi_{i-1}^{\circ}(\bm X_i)
=
\frac{
\left\{\sum_{k<i}K_{\bm h}(\bm X_k-\bm X_i)
\mathbf 1(Z_k^{\circ}=0)\right\}^{2}
}{
\left\{\sum_{k<i}K_{\bm h}(\bm X_k-\bm X_i)
\mathbf 1(Z_k^{\circ}=0)\right\}^{2}
+
\left\{\sum_{k<i}K_{\bm h}(\bm X_k-\bm X_i)
\mathbf 1(Z_k^{\circ}=1)\right\}^{2}
}.
\]
The oracle probability is computed from the covariates and preceding
oracle assignments and is therefore outcome-independent. Let
$\mathcal G_N=\sigma(\mathcal D_H,\bm X_1,\ldots,\bm X_N,
U_1,\ldots,U_N)$.
For $z\in\{0,1\}$, let
$\bm S_{zN}^{\circ}(\bm x)\in\mathbb R^N$ be the normalized kernel row based on
the oracle participants assigned to arm $z$, with zeros outside that arm. Define
$\bm\alpha_{zN}^{\circ}=N^{-1}\sum_{i=1}^N
\bm S_{zN}^{\circ}(\bm X_i)$ for $z\in\{0,1\}$, and let
$\widehat\Delta_N^{\circ}=\sum_i\alpha_{1N,i}^{\circ}Y_i(1)
-\sum_i\alpha_{0N,i}^{\circ}Y_i(0)$.
Both oracle coefficient vectors are $\mathcal G_N$-measurable. Because their
supports are disjoint, exactly one potential outcome from each participant enters the
oracle estimator. Define the oracle smoothing bias as
$B_N^{\circ}=\sum_{i=1}^N\alpha_{1N,i}^{\circ}\mu_1(\bm X_i)
-\sum_{i=1}^N\alpha_{0N,i}^{\circ}\mu_0(\bm X_i)-\Delta_N$.

\phantomsection\label{cond:C3}
\paragraph{(C3) Negligible integrated smoothing bias}
The oracle smoothing bias satisfies
\begin{equation*}
B_N^{\circ}=o_p(N^{-1/2}).
\end{equation*}
This condition concerns the signed average prediction error over the enrolled
trial covariates, which is distinct from the pointwise error of either kernel regression.
Root-$N$ theory for averages of kernel estimators commonly controls this type of
integrated bias by undersmoothing~\citep{Newey1994partial}.  A closely related
precedent within kernel-based adaptive randomization is the second-order-kernel
condition of~\citet{Jiang2018KBCD}, which combines a diverging local-information
rate with a vanishing root-$N$ bias rate.  Lemma
\ref{lem:supp_C3_feasibility} below records exact-reproduction and
undersmoothing regimes that imply \condref{C3}. These regimes provide sufficient conditions for the required integrated-bias control. In the implemented
design, Scott's rule is a prespecified finite-sample bandwidth choice rather
than an asymptotic undersmoothing prescription. With fixed $N_H$, the
implemented Scott bandwidth satisfies \condref{C3} under the exact-reproduction
regime described below.
More generally, \condref{C3} specifies the integrated bias along the trial
sequence, whose verification depends on the response surface and bandwidth
regime.

\phantomsection\label{cond:C4}
\paragraph{(C4) Allocation perturbation, terminal borrowing, and kernel regularity}\leavevmode\par\nobreak
Write $\Lambda_N=\sum_{i=1}^N
|\phi_{i-1}(\bm X_i)-\phi_{i-1}^{\circ}(\bm X_i)|$. For the final analysis,
let $W_{N,i}^{F}=W_N^F(\bm X_i)$ and
$\widetilde n_{0,N,i}^{\mathrm{eff}}=\allowbreak
\max\{n_{0,N}^{\mathrm{eff}}(\bm X_i),n_{\min}\}$,
where $\widehat\sigma_{0,N}^2(\bm X_i)$ is the completed-cohort
variance defined by setting $n=N$ in the Stage~II local summary. Then
\[
\mathrm{SE}_{N,i}^2
=\widehat\tau_H^2(\bm X_i)
+\frac{\widehat\sigma_{0,N}^2(\bm X_i)}{\widetilde n_{0,N,i}^{\mathrm{eff}}},
\qquad
G_{N,i}^{F}
=\frac{\widehat\sigma_{0,N}^2(\bm X_i)}
{\widetilde n_{0,N,i}^{\mathrm{eff}}\mathrm{SE}_{N,i}^2}.
\]
Assume \(\mathrm{SE}_{N,i}^2>0\) for every evaluation profile and that
the completed-cohort marginal discrepancy is well defined with probability
tending to one. The local variance quantities entering denominators or square
roots are floored at $\epsilon_v=10^{-12}$ in
Algorithms~\ref{alg:stage1}--\ref{alg:stage3} in Section~\ref{supp:algorithms}; well-definedness of the marginal
discrepancy is imposed separately. Quantities may be assigned arbitrary finite
values on the exceptional events where the marginal discrepancy is undefined.
There are deterministic sequences $d_N\geq1$ and $r_N\downarrow0$ such that
\begin{equation}
\Lambda_N=O_p(d_N),
\qquad
\frac1N\sum_{i=1}^NG_{N,i}^{F}=O_p(r_N),
\qquad
N^{-1/2+1/q}d_N\to0,
\qquad
N^{1/2+1/q}r_N\to0.
\label{eq:supp_primitive_rates}
\end{equation}
Let $\bm S_{zN}^{C}(\bm X_i)$ denote the normalized concurrent-only kernel row
formed from the actual assignment path, and retain
$\bm S_{zN}^{\circ}(\bm X_i)$ for the oracle path. The subscript $N$ and
superscripts $C$ or $\circ$ denote full-length rows on the common participant index;
the unindexed $\bm S_z$ and $\bm S_H$ in Stage~III are arm-compressed rows used
for computation. Both paths use the same bandwidth specified for the $N$th
trial; this bandwidth is
$\mathcal H_N$-measurable and hence fixed before the auxiliary
randomization is realized. All row
denominators are positive, and the normalized kernel matrices have bounded
column load:
\begin{equation}
\max_{z\in\{0,1\},\,1\leq j\leq N}
\sum_{i=1}^N
\left\{
S_{zN,j}^{C}(\bm X_i)+S_{zN,j}^{\circ}(\bm X_i)
\right\}
=O_p(1).
\label{eq:supp_kernel_regular}
\end{equation}
Entries outside the relevant arm are zero.  Condition
\eqref{eq:supp_kernel_regular} prevents one observed outcome from receiving
unbounded aggregate weight across all evaluation profiles.  It is an average
leverage condition. Its relation to a uniform density lower bound over the
entire covariate support depends on additional kernel and design assumptions.
Together, these requirements state that the
normalized kernel regressions are stable and that the cumulative difference
between the actual and oracle assignment probabilities grows more slowly than
the scale specified in \eqref{eq:supp_primitive_rates}.

The existing KBCD theory establishes balance properties under smooth,
stabilizing allocation maps for outcome-independent kernel biased-coin designs
\citep{Jiang2018KBCD}, and classical biased-coin theory develops the same
feedback principle~\citep{Smith1984properties}. For the outcome-adaptive RADISH
rule, the first relation in \eqref{eq:supp_primitive_rates} summarizes the
required design stability. Lemma~\ref{lem:supp_C4_feasibility} verifies the
remaining rate requirements when this one-over-information coupling rate and
the stated borrowing-rate bound hold.

\phantomsection\label{cond:C5}
\paragraph{(C5) Variance scale and no dominant oracle coefficient}\leavevmode\par\nobreak
Let $V_N^{\circ}=\allowbreak\sigma_1^2\|\bm\alpha_{1N}^{\circ}\|_2^2
+\sigma_0^2\|\bm\alpha_{0N}^{\circ}\|_2^2$.
For some constant $v\in(0,\infty)$,
\begin{equation*}
NV_N^{\circ}\xrightarrow{p}v,
\qquad
\max_{z\in\{0,1\},\,1\leq i\leq N}
\frac{|\alpha_{zN,i}^{\circ}|}{\sqrt{V_N^{\circ}}}
\xrightarrow{p}0.
\end{equation*}
The second relation is a Lindeberg-type condition requiring that no single
coefficient dominates the oracle linear smoother. It is motivated by classical kernel-regression
regimes with diverging local effective sample sizes and covariate density bounded
away from zero on the analysis region
\citep{Hardle1990kernel,WandJones1995kernel}.

\phantomsection\label{cond:C6}
\paragraph{(C6) Oracle residual-smoother regularity}
For $z\in\{0,1\}$, let $\bm A_{zN}^{\circ}$ be the $N\times N$ diagonal matrix
with $i$th diagonal entry $\mathbf 1(Z_i^{\circ}=z)$, and let
$\bm S_{zN}^{\mathrm{self},\circ}$ be the full-length oracle self-smoothing
matrix whose $i$th row equals $\bm S_{zN}^{\circ}(\bm X_i)$ when
$Z_i^{\circ}=z$ and is zero otherwise.  Let
$\bm\mu_{zN}=\{\mu_z(\bm X_1),\ldots,\mu_z(\bm X_N)\}^{\mathsf T}$ and
$N_z^{\circ}=\operatorname{tr}(\bm A_{zN}^{\circ})$.  For constants
$\pi_z\in(0,1)$,
\begin{equation}
\frac{N_z^{\circ}}N\xrightarrow{p}\pi_z,
\qquad
\frac1N\|\bm S_{zN}^{\mathrm{self},\circ}\|_{F}^{2}
\xrightarrow{p}0,
\qquad
\frac1N\|
(\bm A_{zN}^{\circ}-\bm S_{zN}^{\mathrm{self},\circ})
\bm\mu_{zN}\|_2^2
\xrightarrow{p}0.
\label{eq:supp_residual_regular}
\end{equation}
The historical residual degrees of freedom in Section~S4.3 are
positive.  The second requirement in \eqref{eq:supp_residual_regular} is a
vanishing-average-leverage condition, and the third is mean-square consistency
of the oracle mean fit on its assigned arm.  Both follow under familiar kernel
regression conditions such as diverging local effective sample sizes, vanishing
maximum normalized weight, and sufficient smoothness with a shrinking bandwidth
\citep{Andrews1995,MackSilverman1982}. With a fixed nonzero bandwidth, the
third relation holds, for example, when the smoother exactly reproduces each
arm-specific conditional mean at the enrolled covariates. More generally, it
requires the mean-square approximation error to vanish.
The positive limiting arm fractions are the usual positivity requirement and
arise when the oracle biased-coin design converges to an interior allocation
target, as in the balance theory of~\citet{Jiang2018KBCD}.
Condition \condref{C6} is stated in terms of the smoothing operators.
Lemma~\ref{lem:supp_residual_variance_consistency} uses this condition together with the
RADISH-to-oracle stability in \condref{C4} to prove consistency of the actual
residual variance estimators. Under heteroskedastic outcomes, the scalar source-specific
variance estimators in Equation~(7) of the main text would instead need an
observation-level robust replacement.

\begin{lemma}[Sufficient regimes for \condref{C3}]
\label{lem:supp_C3_feasibility}
For $z\in\{0,1\}$, define the oracle fitted conditional mean as
$m_{zN}^{\circ}(\bm X_i)=
\bm S_{zN}^{\circ}(\bm X_i)^{\mathsf T}\bm\mu_{zN}$.
Condition \condref{C3} holds in either of the following regimes:
\begin{enumerate}
\item the oracle smoother reproduces both conditional means at every enrolled
covariate, so that
$m_{zN}^{\circ}(\bm X_i)=\mu_z(\bm X_i)$ for all $i$ and $z$;
\item for some smoothness order $s>0$ and bandwidth vector $\bm h_N$,
\[
\frac1N\sum_{i=1}^N
\{m_{zN}^{\circ}(\bm X_i)-\mu_z(\bm X_i)\}
=O_p(\|\bm h_N\|^s),
\qquad z\in\{0,1\},
\]
and $\sqrt N\|\bm h_N\|^s\to0$.
\end{enumerate}
The first regime includes arm-specific constant conditional means because every
normalized kernel row reproduces constants.  In an isotropic $d$-dimensional
setting with $h_N=N^{-a}$, the usual requirements
$Nh_N^d\to\infty$ and $\sqrt N h_N^s\to0$ can hold simultaneously whenever
$s>d/2$, by choosing $1/(2s)<a<1/d$.  Taking $s=2$, an
integrated-bias expansion therefore permits such a choice for $d<4$.
\end{lemma}

\begin{proof}
First express the smoothing bias in terms of fitted conditional means. For each
$z\in\{0,1\}$, $(\bm\alpha_{zN}^{\circ})^{\mathsf T}\bm\mu_{zN}
=N^{-1}\sum_i\bm S_{zN}^{\circ}(\bm X_i)^{\mathsf T}\bm\mu_{zN}
=N^{-1}\sum_i m_{zN}^{\circ}(\bm X_i)$.
Substituting this equality and the definition of $\Delta_N$ into
$B_N^{\circ}$ gives $B_N^{\circ}=N^{-1}\sum_i
[\{m_{1N}^{\circ}(\bm X_i)-\mu_1(\bm X_i)\}
-\{m_{0N}^{\circ}(\bm X_i)-\mu_0(\bm X_i)\}]$.
In the exact-reproduction regime, both bracketed errors vanish for every $i$,
so $B_N^{\circ}=0$. In the second regime, each signed average is
$O_p(\|\bm h_N\|^s)$. Their difference has the same order by the triangle
inequality, and $\sqrt N\,B_N^{\circ}=O_p(\sqrt N\|\bm h_N\|^s)=o_p(1)$.
This is equivalent to $B_N^{\circ}=o_p(N^{-1/2})$.

The two bandwidth requirements are compatible. If $h_N=N^{-a}$, then
$Nh_N^d=N^{1-ad}\to\infty$ exactly when $a<1/d$, whereas
$\sqrt N h_N^s=N^{1/2-as}\to0$ exactly when $a>1/(2s)$. A value of $a$ in
this interval exists if and only if $1/(2s)<1/d$, or equivalently $s>d/2$.
\end{proof}

\begin{lemma}[A sufficient one-over-information regime for \condref{C4}]
\label{lem:supp_C4_feasibility}\leavevmode\par\nobreak
Suppose, in addition to the kernel-row requirement
\eqref{eq:supp_kernel_regular}, that
\[
\sup_{i_0\leq i\leq N}
i\,|\phi_{i-1}(\bm X_i)-\phi_{i-1}^{\circ}(\bm X_i)|
=O_p(1),
\qquad
\max_{1\leq i\leq N}NG_{N,i}^{F}=O_p(1),
\]
where $i_0$ is the fixed end of burn-in.  Then the rate requirements in
\eqref{eq:supp_primitive_rates} hold with
$d_N=\log(eN)$ and $r_N=N^{-1}$ for every $q>2$.
\end{lemma}

\begin{proof}
Define $A_N=\sup_{i_0\leq i\leq N}
i\,|\phi_{i-1}(\bm X_i)-\phi_{i-1}^{\circ}(\bm X_i)|$.
The two designs share the burn-in, so their allocation probabilities agree
before $i_0$. For $i\geq i_0$, the definition of $A_N$ gives a pointwise bound
by $A_N/i$. Hence $\Lambda_N\leq A_N\sum_{i=i_0}^Ni^{-1}
\leq A_N\{1+\log N\}=O_p\{\log(eN)\}$.
For the borrowing term, the second premise of the lemma gives
$\max_i N G_{N,i}^{F}=O_p(1)$. The mean borrowing weight is bounded by its
maximum, and hence $N^{-1}\sum_iG_{N,i}^{F}=O_p(N^{-1})$.
Thus $d_N=\log(eN)$ and $r_N=N^{-1}$ provide the two stochastic bounds in
\eqref{eq:supp_primitive_rates}. Since $q>2$, the exponent
$-1/2+1/q$ is negative. Therefore
$N^{-1/2+1/q}\log(eN)\to0$ and
$N^{1/2+1/q}N^{-1}=N^{-1/2+1/q}\to0$.
This verifies the remaining deterministic rate restrictions.
\end{proof}

The preceding regime has a direct design interpretation. If local concurrent
information in both arms grows proportionally to enrollment, the fixed
historical reference changes the effective local control information by order
one, whereas the allocation totals are of order $i$. At fixed local totals, a
smooth allocation rule then changes the allocation probability by order
$i^{-1}$. The preceding lemma assumes this uniform rate for the full coupled
outcome-adaptive recursion, while condition~\condref{C4} allows the more general
cumulative perturbation bound.
Similarly, local control information of order $N$, a bounded
concurrent variance estimate, and a historical reference variance bounded away
from zero give $G_{N,i}^{F}=O_p(N^{-1})$.

\subsection{Oracle Equivalence}
\label{supp:oracle_equivalence}

For $z\in\{0,1\}$, let $\widehat{\bm\alpha}_{zN}^{C}$ be the full-length
concurrent-only coefficient vector obtained by averaging
$\bm S_{zN}^{C}(\bm X_i)$ over $i=1,\ldots,N$, and let
$\widehat\Delta_N^{C}=\sum_i\widehat\alpha_{1N,i}^{C}Y_i(1)
-\sum_i\widehat\alpha_{0N,i}^{C}Y_i(0)$.
This intermediate estimator uses the actual RADISH assignment path but sets the
terminal borrowing weights to zero. It separates the effect of terminal
borrowing from the effect of the outcome-dependent allocation path.

\begin{lemma}[Normalization of two kernel rows]
\label{lem:supp_row_normalization}
\begin{sloppypar}
Let $k_1,\ldots,k_N$ be nonnegative, let $A$ and $B$ be two nonempty index sets,
and define
$p_{A,j}=k_j\mathbf 1(j\in A)/s_A$ and
$p_{B,j}=k_j\mathbf 1(j\in B)/s_B$, where
$s_A=\sum_{j\in A}k_j>0$ and $s_B=\sum_{j\in B}k_j>0$. Then
\end{sloppypar}
\[
\|\bm p_A-\bm p_B\|_1
\leq
2\left(
\sum_{j\in A\setminus B}p_{A,j}
+
\sum_{j\in B\setminus A}p_{B,j}
\right).
\]
\end{lemma}

\begin{proof}
Let $u_A=\sum_{j\in A\setminus B}p_{A,j}$ and
$u_B=\sum_{j\in B\setminus A}p_{B,j}$. The indices outside $A\cap B$
contribute exactly $u_A+u_B$ to the $\ell_1$ distance. On $A\cap B$,
$p_{A,j}-p_{B,j}=k_j(s_A^{-1}-s_B^{-1})$, so all nonzero differences have the
same sign. Their absolute values can therefore be summed before taking the
absolute value, yielding $\sum_{j\in A\cap B}|p_{A,j}-p_{B,j}|
=|\sum_{j\in A\cap B}p_{A,j}-\sum_{j\in A\cap B}p_{B,j}|=|u_A-u_B|$.
Consequently,
$\|\bm p_A-\bm p_B\|_1=u_A+u_B+|u_A-u_B|\leq2(u_A+u_B)$, as claimed.
\end{proof}

\begin{lemma}[First-order equivalence of RADISH and its oracle]
\label{lem:supp_oracle_equivalence}
Under \condref{C1}, \condref{C2}, and \condref{C4},
\begin{equation}
\widehat\Delta_N-\widehat\Delta_N^{\circ}=o_p(N^{-1/2}),
\label{eq:supp_point_stability}
\end{equation}
and, on defining $\bm\alpha_{HN}^{\circ}$ as the zero vector in
$\mathbb R^{N_H}$,
\begin{equation}
\sum_{z\in\{1,0,H\}}
\|\widehat{\bm\alpha}_{zN}-\bm\alpha_{zN}^{\circ}\|_2
=o_p(N^{-1/2}).
\label{eq:supp_coefficient_stability}
\end{equation}
\end{lemma}

\begin{proof}
To compare the allocation paths, let $I_i=\mathbf 1(Z_i\ne Z_i^{\circ})$ and
$J_N=\sum_{i=1}^NI_i$.  Let $\mathcal H_{i-1}^{X}$ be the joint information
generated by the historical data, the covariates observed through participant $i$,
the preceding actual outcomes and assignments, and the preceding oracle
assignments. Because the two paths use the same $U_i$,
$\E(I_i\mid\mathcal H_{i-1}^{X})=
|\phi_{i-1}(\bm X_i)-\phi_{i-1}^{\circ}(\bm X_i)|$.
Let $p_{N,i}=|\phi_{i-1}(\bm X_i)-
\phi_{i-1}^{\circ}(\bm X_i)|$ and
$\mathcal M_{N,k}=\sum_{i=1}^k(I_i-p_{N,i})$. Its increment is
$\Delta\mathcal M_{N,i}=I_i-p_{N,i}$. Because $I_i$ is conditionally Bernoulli with
success probability $p_{N,i}$,
$\E(\Delta\mathcal M_{N,i}\mid\mathcal H_{i-1}^{X})=0$ and
$\E\{(\Delta\mathcal M_{N,i})^2\mid\mathcal H_{i-1}^{X}\}
=p_{N,i}(1-p_{N,i})$.
Thus $\{\mathcal M_{N,k}:0\leq k\leq N\}$ is a square-integrable martingale. Its
predictable quadratic variation is, by definition,
\[
\langle \mathcal M_N\rangle_k
:=
\sum_{i=1}^k
\E\{(\Delta\mathcal M_{N,i})^2\mid\mathcal H_{i-1}^{X}\}
=
\sum_{i=1}^k p_{N,i}(1-p_{N,i})
\leq
\sum_{i=1}^k p_{N,i}.
\]
Fix $c_{\mathrm{stop}}>0$, and let $\tau_N$ be the first $k\leq N$ for which
$\sum_{i=1}^k p_{N,i}>c_{\mathrm{stop}}d_N$, with $\tau_N=N$ when no crossing occurs. Immediately
before a crossing, the cumulative sum is at most $c_{\mathrm{stop}}d_N$, and the crossing
increment is at most one. Consequently,
$\langle \mathcal M_N\rangle_{N\wedge\tau_N}
\leq c_{\mathrm{stop}}d_N+1$.
Because $p_{N,k}$ is measurable before the $k$th randomization,
$\tau_N$ is a bounded stopping time for the associated martingale filtration.
Let $\mathcal E_N(c_{\mathrm{stop}})=\{\Lambda_N\leq c_{\mathrm{stop}}d_N\}$. No crossing occurs on $\mathcal E_N(c_{\mathrm{stop}})$, so
$\mathcal M_{N,N}=\mathcal M_{N,N\wedge\tau_N}$ there. For every $L>0$, split the probability
according to $\mathcal E_N(c_{\mathrm{stop}})$:
\[
\begin{aligned}
\Pr\{|\mathcal M_{N,N}|>Ld_N\}
&\leq
\Pr\{\mathcal E_N(c_{\mathrm{stop}})^c\}
+
\Pr\{|\mathcal M_{N,N}|>Ld_N,\ \mathcal E_N(c_{\mathrm{stop}})\}\\
&\leq
\Pr(\Lambda_N>c_{\mathrm{stop}}d_N)
+
\Pr\{|\mathcal M_{N,N\wedge\tau_N}|>Ld_N\}\\
&\leq
\Pr(\Lambda_N>c_{\mathrm{stop}}d_N)
+
\frac{\E(\mathcal M_{N,N\wedge\tau_N}^2)}{L^2d_N^2}\\
&=
\Pr(\Lambda_N>c_{\mathrm{stop}}d_N)
+
\frac{\E\{\langle \mathcal M_N\rangle_{N\wedge\tau_N}\}}
{L^2d_N^2}\\
&\leq
\Pr(\Lambda_N>c_{\mathrm{stop}}d_N)
+
\frac{c_{\mathrm{stop}}d_N+1}{L^2d_N^2}\\
&\leq
\Pr(\Lambda_N>c_{\mathrm{stop}}d_N)
+
\frac{c_{\mathrm{stop}}+1}{L^2d_N}.
\end{aligned}
\]
The third line is Chebyshev's inequality. The equality on the fourth line is
the martingale isometry for the stopped square-integrable martingale. For any
$\epsilon>0$, $\Lambda_N=O_p(d_N)$ permits a fixed $c_{\mathrm{stop}}$ for which
the first term is below $\epsilon/2$ for all sufficiently large $N$. With this
$c_{\mathrm{stop}}$ fixed, choose $L$ so that
$(c_{\mathrm{stop}}+1)/L^2<\epsilon/2$; the second term is then below
$\epsilon/2$ because $d_N\geq1$. This proves
$\mathcal M_{N,N}/d_N=O_p(1)$. Finally,
$J_N=\sum_{i=1}^NI_i=\sum_{i=1}^Np_{N,i}+\mathcal M_{N,N}
=\Lambda_N+\mathcal M_{N,N}$, so $J_N=O_p(d_N)$. Hence
\begin{equation}
J_N=O_p(d_N).
\label{eq:supp_mismatch_rate}
\end{equation}

The mismatch bound also controls the difference between the averaged kernel
coefficients.
Let $\mathcal J_N=\{j:Z_j\ne Z_j^{\circ}\}$, so
$|\mathcal J_N|=J_N$. For a fixed evaluation participant $i$ and arm $z$, the
actual and oracle kernel rows use the same nonnegative kernel values. Their
supports can differ only at indices in $\mathcal J_N$. Lemma
\ref{lem:supp_row_normalization} therefore gives
\[
\|\bm S_{zN}^{C}(\bm X_i)-\bm S_{zN}^{\circ}(\bm X_i)\|_1
\leq
2\sum_{j\in\mathcal J_N}
\{S_{zN,j}^{C}(\bm X_i)+S_{zN,j}^{\circ}(\bm X_i)\}.
\]
The coefficient vectors are averages of these rows. The triangle inequality,
followed by a change in the order of summation, now gives
\begin{equation}
\begin{aligned}
\|\widehat{\bm\alpha}_{zN}^{C}-\bm\alpha_{zN}^{\circ}\|_1
&\leq
\frac1N\sum_{i=1}^N
\|\bm S_{zN}^{C}(\bm X_i)-\bm S_{zN}^{\circ}(\bm X_i)\|_1\\
&\leq
\frac2N\sum_{j:Z_j\ne Z_j^{\circ}}
\sum_{i=1}^N
\{S_{zN,j}^{C}(\bm X_i)+S_{zN,j}^{\circ}(\bm X_i)\}\\
&=O_p(J_N/N)
=O_p(d_N/N),
\qquad z\in\{0,1\}.
\end{aligned}
\label{eq:supp_path_coefficient_rate}
\end{equation}
The penultimate equality uses the maximum column-load bound in
\eqref{eq:supp_kernel_regular}, and the final equality uses
\eqref{eq:supp_mismatch_rate}.
The same bound holds for the Euclidean norm because
$\|\bm v\|_2\leq\|\bm v\|_1$.

To transfer the coefficient bound to the estimators, we control the largest
concurrent outcome.
For any $c_{\mathrm{tail}}>0$, the union bound and the uniform $q$th-moment bound in
\condref{C1} give
\[
\Pr\!\left\{\max_{i,z}|Y_i(z)|>c_{\mathrm{tail}} N^{1/q}\right\}
\leq
c_{\mathrm{tail}}^{-q}N^{-1}\sum_{i=1}^N\sum_{z=0}^1\E|Y_i(z)|^q
\leq 2C c_{\mathrm{tail}}^{-q},
\]
where the first inequality is Markov's inequality applied after the union
bound, and the second follows by taking expectations in the conditional moment
bound. Letting $c_{\mathrm{tail}}$ increase proves
\begin{equation}
\max_{i,z}|Y_i(z)|=O_p(N^{1/q}).
\label{eq:supp_max_outcome}
\end{equation}
Combining \eqref{eq:supp_path_coefficient_rate} and
\eqref{eq:supp_max_outcome} yields
$|\widehat\Delta_N^{C}-\widehat\Delta_N^{\circ}|
=O_p(d_NN^{-1+1/q})=o_p(N^{-1/2})$ by the rate imposed on $d_N$ in
\eqref{eq:supp_primitive_rates}.

It remains to bound the contribution of terminal historical borrowing.
Since $D_N^F(\bm X_i)\geq0$, the Stage~III definition and
$W_{N,i}^F=(R_{N,i}^F-1)/R_{N,i}^F$ give
\[
0\leq W_{N,i}^{F}
\leq R_{N,i}^F-1
\leq
\frac{\exp\{-D_N^F(\bm X_i)\}\,\widehat\sigma_{0,N}^2(\bm X_i)}
 {\widetilde n_{0,N,i}^{\mathrm{eff}}\mathrm{SE}_{N,i}^2}
\leq
\frac{\widehat\sigma_{0,N}^2(\bm X_i)}
 {\widetilde n_{0,N,i}^{\mathrm{eff}}\mathrm{SE}_{N,i}^2}
=G_{N,i}^{F}.
\]
Therefore $\overline W_N^{F}:=N^{-1}\sum_{i=1}^NW_{N,i}^{F}=O_p(r_N)$.
The historical fit $\theta(\bm x)$ and concurrent-control fit
$\bar Y_{0,N}(\bm x)$ are the kernel fits used at final analysis.
The historical sample is fixed after conditioning on $\mathcal D_H$.
Both fits are convex combinations,
and hence
\[
\max_{1\leq i\leq N}
|\theta(\bm X_i)-\bar Y_{0,N}(\bm X_i)|
\leq \max_{1\leq j\leq N_H}|Y_j^H|+\max_{i,z}|Y_i(z)|
=O_p(N^{1/q}).
\]
The exact identity
\[
\widehat\Delta_N-\widehat\Delta_N^{C}
=
-\frac1N\sum_{i=1}^NW_{N,i}^{F}
\{\theta(\bm X_i)-\bar Y_{0,N}(\bm X_i)\}
\]
and \eqref{eq:supp_max_outcome} consequently give
$|\widehat\Delta_N-\widehat\Delta_N^{C}|
=O_p(r_NN^{1/q})=o_p(N^{-1/2})$.
Nonnegativity and unit row sums also give
\[
\|\widehat{\bm\alpha}_{HN}\|_1=\overline W_N^{F},
\qquad
\|\widehat{\bm\alpha}_{0N}-\widehat{\bm\alpha}_{0N}^{C}\|_1
=\overline W_N^{F},
\qquad
\widehat{\bm\alpha}_{1N}=\widehat{\bm\alpha}_{1N}^{C}.
\]
The rate for $r_N$ in \eqref{eq:supp_primitive_rates} implies
$r_N=o(N^{-1/2})$, so each displayed $\ell_1$ norm, and therefore each
corresponding $\ell_2$ norm, is $o_p(N^{-1/2})$.

Combining the allocation and terminal-borrowing bounds by the triangle
inequality proves \eqref{eq:supp_point_stability}. Combining
\eqref{eq:supp_path_coefficient_rate} with the terminal coefficient bounds
proves \eqref{eq:supp_coefficient_stability}, because
\[
\frac{d_N/N}{N^{-1/2}}
=d_NN^{-1/2}
=\{d_NN^{-1/2+1/q}\}N^{-1/q}\to0,
\qquad r_N=o(N^{-1/2}).
\]
\end{proof}

\begin{proof}[Proof of Proposition~\ref{prop:radish_design_reversion}]
\begin{sloppypar}
For the assignment limit, \eqref{eq:supp_mismatch_rate} gives
$J_N=\sum_{i=1}^N\mathbf 1(Z_i\ne Z_i^{\circ})=O_p(d_N)$. Moreover,
$d_N/N=\allowbreak\{d_NN^{-1/2+1/q}\}N^{-1/2-1/q}\allowbreak\to0$ by
\eqref{eq:supp_primitive_rates}. Hence
$J_N/N=O_p(d_N/N)\xrightarrow{p}0$.
\end{sloppypar}

For the borrowing limit, the terminal-weight bound in the proof of
Lemma~\ref{lem:supp_oracle_equivalence} gives
\[
N^{-1}\sum_{i=1}^NW_N^F(\bm X_i)
=\overline W_N^F
=O_p(r_N).
\]
Because $r_N\to0$, this average converges in probability to zero. Finally,
\eqref{eq:supp_point_stability} is exactly the asserted first-order estimator
equivalence.
\end{proof}

\subsection{Oracle Central Limit Theorem}
\label{supp:oracle_clt}

\begin{lemma}[Conditional weighted central limit theorem]
\label{lem:supp_oracle_clt}
Under \condref{C1}, \condref{C2}, and \condref{C5},
\[
\frac{
\widehat\Delta_N^{\circ}-\Delta_N-B_N^{\circ}
}{\sqrt{V_N^{\circ}}}
\xrightarrow{d}\Normal(0,1).
\]
\end{lemma}

\begin{proof}
Conditioning on the auxiliary randomization variables yields a sum of
independent errors. For participant $i$, define
$\psi_{N,i}=\alpha_{1N,i}^{\circ}\varepsilon_i(1)
-\alpha_{0N,i}^{\circ}\varepsilon_i(0)$.
The two oracle coefficient vectors have disjoint supports, so at most one term
in $\psi_{N,i}$ is nonzero. Conditioning on $\mathcal G_N$ fixes the enrolled
covariates, the oracle assignments, and the oracle coefficients, while leaving
the potential-outcome errors unobserved because the auxiliary uniforms are
independent of those errors under \condref{C2}. The cross-participant independence
and conditional mean-zero assumptions in \condref{C1} therefore give
$\E(\psi_{N,i}\mid\mathcal G_N)=0$, with
$\psi_{N,1},\ldots,\psi_{N,N}$ conditionally independent.
Source-wise homoskedasticity and disjoint support then imply
\[
\begin{aligned}
\sum_{i=1}^N\Var(\psi_{N,i}\mid\mathcal G_N)
&=
\sigma_1^2\sum_{i=1}^N(\alpha_{1N,i}^{\circ})^2
+
\sigma_0^2\sum_{i=1}^N(\alpha_{0N,i}^{\circ})^2\\
&=V_N^{\circ}.
\end{aligned}
\]

For the Lyapunov condition, let
$\underline\sigma^2=\min(\sigma_0^2,\sigma_1^2)>0$ and
$a_N=\max_{z,i}|\alpha_{zN,i}^{\circ}|/\sqrt{V_N^{\circ}}$.
The conditional $q$th-moment bound in \condref{C1} gives a finite constant
$C_\varepsilon$ such that
$\E(|\psi_{N,i}|^q\mid\mathcal G_N)
\leq C_\varepsilon\sum_{z=0}^1|\alpha_{zN,i}^{\circ}|^q$.
After summing over $i$ and factoring each coefficient power into a maximum
factor of order $q-2$ and a squared coefficient, we obtain
\begin{align*}
\frac{\sum_{i=1}^N\E(|\psi_{N,i}|^q\mid\mathcal G_N)}
{(V_N^{\circ})^{q/2}}
&\leq
C_\varepsilon a_N^{q-2}
\frac{\sum_{z=0}^1\sum_{i=1}^N
 (\alpha_{zN,i}^{\circ})^2}{V_N^{\circ}}\\
&\leq
\frac{C_\varepsilon}{\underline\sigma^2}a_N^{q-2}
=o_p(1),
\end{align*}
because
$V_N^{\circ}\geq\underline\sigma^2
\sum_{z,i}(\alpha_{zN,i}^{\circ})^2$ and $a_N\xrightarrow{p}0$ by
\condref{C5}. This is the conditional Lyapunov ratio.

To average over the auxiliary randomization variables while retaining the base
conditioning, take an arbitrary subsequence of trial sizes. Convergence in
probability of the Lyapunov ratio yields a further subsequence on which that
ratio converges to zero almost surely. Along this further subsequence, the conditional
Lyapunov theorem applies to the corresponding triangular array and gives, for
almost every realization of the conditioning sequence,
\[
\E\left[
\left.
\exp\left\{it\frac{\sum_i\psi_{N,i}}{\sqrt{V_N^{\circ}}}\right\}
\right|\mathcal G_N
\right]
\longrightarrow
\exp(-t^2/2)
\]
for every fixed $t\in\mathbb R$. The conditional characteristic functions
have absolute value at most one. Dominated convergence, after averaging over
the auxiliary uniforms under the fixed base conditioning, yields the desired
conditional characteristic-function convergence. Every subsequence has a further subsequence
with this same limit, so the subsequence principle gives
\[
\frac{\sum_{i=1}^N\psi_{N,i}}{\sqrt{V_N^{\circ}}}
\xrightarrow{d}\Normal(0,1)
\]
for the original sequence. Because the statistic remains centered at the
conditioned target $\Delta_N$, the limiting variance remains conditional on the
enrolled covariates. Finally, expanding the oracle estimator and
using the definition of $B_N^{\circ}$ gives
$\widehat\Delta_N^{\circ}-\Delta_N-B_N^{\circ}=\sum_i\psi_{N,i}$.
\end{proof}

\subsection{Consistency of the Working Variance}
\label{supp:variance_consistency}

\begin{lemma}[Consistency of the concurrent residual variance estimators]
\label{lem:supp_residual_variance_consistency}\leavevmode\par\nobreak
Under \condref{C1}, \condref{C2}, \condref{C4}, and \condref{C6},
\[
\widehat\sigma_z^2\xrightarrow{p}\sigma_z^2,
\qquad z\in\{0,1\},
\]
and $\widehat\sigma_H^2=O_p(1)$.
\end{lemma}

\begin{proof}
Fix $z\in\{0,1\}$.
Define the full-length vectors
$\bm Y_N(z)=\{Y_1(z),\ldots,Y_N(z)\}^{\mathsf T}$ and
$\bm\varepsilon_N(z)=\{\varepsilon_1(z),\ldots,\varepsilon_N(z)\}^{\mathsf T}$.

To compare the actual and oracle residual operators, let
$\bm A_{zN}^{C}$ be diagonal with $i$th entry
$\mathbf 1(Z_i=z)$, and let $\bm S_{zN}^{\mathrm{self},C}$ be the
full-length actual self-smoothing matrix defined in the same way as its oracle
counterpart in \condref{C6}. Set
$\bm R_{zN}^{C}=\bm A_{zN}^{C}-\bm S_{zN}^{\mathrm{self},C}$ and
$\bm R_{zN}^{\circ}=\bm A_{zN}^{\circ}-\bm S_{zN}^{\mathrm{self},\circ}$.
Equation~\eqref{eq:supp_mismatch_rate} shows that the number of assignment
mismatches satisfies $J_N=O_p(d_N)$. On a row whose arm indicators agree,
Lemma~\ref{lem:supp_row_normalization} bounds the difference between the two
normalized kernel rows.  On a mismatched row, the $\ell_1$ distance between the
two self-smoothing rows is at most two. Let
$\mathcal J_N=\{j:Z_j\ne Z_j^{\circ}\}$ and let $L_N$ denote the maximum
column load on the left side of \eqref{eq:supp_kernel_regular}. Thus
$L_N=O_p(1)$. Summing first over rows and then over the mismatched columns gives
\begin{equation}
\begin{aligned}
\sum_{i=1}^N
\|\bm S_{zN,i}^{\mathrm{self},C}
-\bm S_{zN,i}^{\mathrm{self},\circ}\|_1
&\leq
2J_N+2\sum_{j\in\mathcal J_N}\sum_{i=1}^N
\{S_{zN,j}^{C}(\bm X_i)+S_{zN,j}^{\circ}(\bm X_i)\}\\
&\leq 2J_N(1+L_N)
=O_p(d_N).
\end{aligned}
\label{eq:supp_self_row_transfer}
\end{equation}
Here the matched rows contribute at most twice the total kernel mass assigned
to columns in $\mathcal J_N$, and the at most $J_N$ mismatched rows each
contribute at most two. Every row difference has $\ell_1$ norm at most two.
Therefore
\[
\begin{aligned}
\|\bm S_{zN}^{\mathrm{self},C}
-\bm S_{zN}^{\mathrm{self},\circ}\|_F^2
&=
\sum_{i=1}^N
\|\bm S_{zN,i}^{\mathrm{self},C}
-\bm S_{zN,i}^{\mathrm{self},\circ}\|_2^2\\
&\leq
2\sum_{i=1}^N
\|\bm S_{zN,i}^{\mathrm{self},C}
-\bm S_{zN,i}^{\mathrm{self},\circ}\|_1
=O_p(d_N).
\end{aligned}
\]

Let $Y_N^{\max}=\max_{i,z}|Y_i(z)|$.  Equation
\eqref{eq:supp_max_outcome} gives $Y_N^{\max}=O_p(N^{1/q})$.  The selector matrices
differ on at most $J_N$ diagonal entries, so
$\|(\bm A_{zN}^{C}-\bm A_{zN}^{\circ})\bm Y_N(z)\|_2^2
\leq J_N(Y_N^{\max})^2$.
For the smoothing part, the rowwise bound
$|\bm v^{\mathsf T}\bm Y_N(z)|\leq Y_N^{\max}\|\bm v\|_1$ and
\eqref{eq:supp_self_row_transfer} give
\[
\|(
\bm S_{zN}^{\mathrm{self},C}
-\bm S_{zN}^{\mathrm{self},\circ})\bm Y_N(z)\|_2^2
\leq
2(Y_N^{\max})^2\sum_{i=1}^N
\|\bm S_{zN,i}^{\mathrm{self},C}
-\bm S_{zN,i}^{\mathrm{self},\circ}\|_1
=O_p\{(Y_N^{\max})^2d_N\}.
\]
Applying $\|\bm a+\bm b\|_2^2\leq
2\|\bm a\|_2^2+2\|\bm b\|_2^2$ to the selector and smoothing parts yields
\begin{align*}
\frac1N\|
(\bm R_{zN}^{C}-\bm R_{zN}^{\circ})\bm Y_N(z)
\|_2^2
&=O_p\left\{\frac{(Y_N^{\max})^2d_N}{N}\right\}\\
&=O_p(d_NN^{-1+2/q})
=o_p(1).
\end{align*}
For the last equality, write the middle rate as
$\{d_NN^{-1/2+1/q}\}N^{-1/2+1/q}$ and use $q>2$ together with
\eqref{eq:supp_primitive_rates}.

For the oracle residual sum of squares, write
$\bm Y_N(z)=\bm\mu_{zN}+\bm\varepsilon_N(z)$. Conditional on
$\mathcal G_N$, the oracle selector and smoother are fixed and independent of
the potential-outcome errors. Let $p=\min(q/2,2)>1$ and define
$\zeta_{N,i}=\mathbf 1(Z_i^{\circ}=z)
\{\varepsilon_i(z)^2-\sigma_z^2\}$.
The variables $\zeta_{N,i}$ are conditionally independent and centered. Since
$2p\leq q$, \condref{C1} also gives
$\sup_i\E(|\zeta_{N,i}|^p\mid\mathcal G_N)=O(1)$ almost surely. The
von Bahr--Esseen inequality~\citep{vonBahrEsseen1965}, with the ordinary
variance identity when $p=2$, therefore yields
\[
\begin{aligned}
\E\left(
\left.
\left|\frac1N\sum_{i=1}^N\zeta_{N,i}\right|^p
\right|\mathcal G_N
\right)
&\leq
\frac{C_p}{N^p}
\sum_{i=1}^N\E(|\zeta_{N,i}|^p\mid\mathcal G_N)\\
&=O(N^{1-p})
\longrightarrow0.
\end{aligned}
\]
Markov's inequality, followed by averaging over $\mathcal G_N$, proves
$N^{-1}\sum_i\zeta_{N,i}\xrightarrow{p}0$. The sum of squared errors for
participants assigned to arm $z$
equals $N_z^{\circ}\sigma_z^2+\sum_i\zeta_{N,i}$. Therefore, the
arm-fraction condition in \condref{C6} gives
$N^{-1}\|\bm A_{zN}^{\circ}\bm\varepsilon_N(z)\|_2^2
\xrightarrow{p}\pi_z\sigma_z^2$.
Next, conditional homoskedasticity and independence give
\[
\E\!\left\{
\|\bm S_{zN}^{\mathrm{self},\circ}\bm\varepsilon_N(z)\|_2^2
\mid\mathcal G_N
\right\}
=
\sigma_z^2\|\bm S_{zN}^{\mathrm{self},\circ}\|_F^2
=o_p(N).
\]
For the conditional Markov argument, set
$Q_N=N^{-1}\|\bm S_{zN}^{\mathrm{self},\circ}
\bm\varepsilon_N(z)\|_2^2$. For any $\epsilon,\epsilon_0>0$,
$\Pr(Q_N>\epsilon)\leq
\Pr\{\E(Q_N\mid\mathcal G_N)>\epsilon_0\}+\epsilon_0/\epsilon$.
The first term tends to zero by \condref{C6}. Letting $\epsilon_0$ tend to zero
after taking the limit shows that $Q_N\xrightarrow{p}0$.

Since $\bm R_{zN}^{\circ}=\bm A_{zN}^{\circ}-
\bm S_{zN}^{\mathrm{self},\circ}$, the difference-of-squares identity and
Cauchy--Schwarz give
\[
\begin{aligned}
&\left|
\frac1N\|\bm R_{zN}^{\circ}\bm\varepsilon_N(z)\|_2^2
-
\frac1N\|\bm A_{zN}^{\circ}\bm\varepsilon_N(z)\|_2^2
\right|\\
&\quad\leq
Q_N
+2Q_N^{1/2}
\left\{
\frac1N\|\bm A_{zN}^{\circ}\bm\varepsilon_N(z)\|_2^2
\right\}^{1/2}
=o_p(1).
\end{aligned}
\]
It follows that $N^{-1}\|\bm R_{zN}^{\circ}\bm\varepsilon_N(z)\|_2^2
\xrightarrow{p}\pi_z\sigma_z^2$.
The mean-approximation requirement in \eqref{eq:supp_residual_regular} gives
$N^{-1}\|\bm R_{zN}^{\circ}\bm\mu_{zN}\|_2^2=o_p(1)$.
Applying the same difference-of-squares bound to
$\bm R_{zN}^{\circ}\bm Y_N(z)=
\bm R_{zN}^{\circ}\bm\mu_{zN}+
\bm R_{zN}^{\circ}\bm\varepsilon_N(z)$ shows that the mean term and its cross
product are $o_p(1)$ after division by $N$. Therefore
\begin{equation}
\frac1N\|\bm R_{zN}^{\circ}\bm Y_N(z)\|_2^2
\xrightarrow{p}\pi_z\sigma_z^2.
\label{eq:supp_oracle_rss_limit}
\end{equation}

The residual-operator bound transfers this limit to the actual RADISH path.
Define
\[
\bm a_N=\bm R_{zN}^{\circ}\bm Y_N(z),
\qquad
\bm b_N=(\bm R_{zN}^{C}-\bm R_{zN}^{\circ})\bm Y_N(z).
\]
The residual-operator bound above and
\eqref{eq:supp_oracle_rss_limit} give, respectively,
\[
N^{-1/2}\|\bm b_N\|_2=o_p(1),
\qquad
N^{-1/2}\|\bm a_N\|_2=O_p(1).
\]
Since $\bm R_{zN}^{C}\bm Y_N(z)=\bm a_N+\bm b_N$,
\[
\left|
\frac1N\|\bm a_N+\bm b_N\|_2^2
-
\frac1N\|\bm a_N\|_2^2
\right|
\leq
\frac{2}{N}\|\bm a_N\|_2\|\bm b_N\|_2
+
\frac1N\|\bm b_N\|_2^2
=o_p(1).
\]
Combining this bound with \eqref{eq:supp_oracle_rss_limit} yields
\begin{equation}
\frac1N\|\bm R_{zN}^{C}\bm Y_N(z)\|_2^2
\xrightarrow{p}\pi_z\sigma_z^2.
\label{eq:supp_actual_rss_limit}
\end{equation}
The numerator on the left is exactly the within-arm residual sum of squares in
Section~S4.3, written on the common participant index.

To obtain the variance limit, we also normalize the residual degrees of
freedom.
Because $|N_z-N_z^{\circ}|\leq J_N$ and $d_N/N\to0$,
$N_z/N\xrightarrow{p}\pi_z$. The Frobenius bound established above and
\eqref{eq:supp_residual_regular} give, by the squared triangle inequality,
\[
\frac1N\|\bm S_{zN}^{\mathrm{self},C}\|_F^2
\leq
\frac2N\|\bm S_{zN}^{\mathrm{self},\circ}\|_F^2
+
\frac2N\|\bm S_{zN}^{\mathrm{self},C}
-\bm S_{zN}^{\mathrm{self},\circ}\|_F^2
\xrightarrow{p}0.
\]
Also,
$|\operatorname{tr}(\bm S_{zN}^{\mathrm{self},C})|
\leq\sqrt N\|\bm S_{zN}^{\mathrm{self},C}\|_F=o_p(N)$.
Let $\operatorname{dof}_{z}^{\mathrm{raw}}$ denote the unfloored
residual degrees of freedom. Writing it on the common participant index gives
\[
\frac{\operatorname{dof}_{z}^{\mathrm{raw}}}{N}
=
\frac{N_z}{N}
-
\frac{2}{N}\operatorname{tr}(\bm S_{zN}^{\mathrm{self},C})
+
\frac1N\|\bm S_{zN}^{\mathrm{self},C}\|_F^2
\xrightarrow{p}\pi_z.
\]
Because $\pi_z>0$, $\Pr\{\operatorname{dof}_{z}^{\mathrm{raw}}>1\}\to1$.
Hence $\operatorname{dof}_z=\max\{\operatorname{dof}_{z}^{\mathrm{raw}},1\}$
has the same normalized limit.
Dividing \eqref{eq:supp_actual_rss_limit} by this positive limit proves
$\widehat\sigma_z^2\xrightarrow{p}\sigma_z^2$. Finally, because each historical
smoothing row is convex and the residual degrees of freedom are bounded below
by one, the historical residual sum of squares is bounded by
$4N_H\max_j|Y_j^H|^2$. Conditional on the fixed historical dataset, this gives
$\widehat\sigma_H^2=O_p(1)$ even when the bandwidth varies along the trial
sequence.
\end{proof}

\begin{lemma}[Working-variance consistency]
\label{lem:supp_variance_consistency}
Under \condref{C1}, \condref{C2}, \condref{C4}, \condref{C5}, and
\condref{C6},
\begin{equation}
\frac{\widehat V_{\mathrm{lin}}}{V_N^{\circ}}
\xrightarrow{p}1,
\qquad
\frac{\widehat V_C}{V_N^{\circ}}
\xrightarrow{p}1,
\qquad
\frac{\widehat V_N}{V_N^{\circ}}
\xrightarrow{p}1.
\label{eq:supp_variance_ratio}
\end{equation}
\end{lemma}

\begin{proof}
Set $\bm\alpha_{HN}^{\circ}=\bm0$ and
$\bm d_{zN}=\widehat{\bm\alpha}_{zN}-\bm\alpha_{zN}^{\circ}$ for
$z\in\{1,0,H\}$.  By \eqref{eq:supp_coefficient_stability},
$\|\bm d_{zN}\|_2=o_p(N^{-1/2})$.  Condition \condref{C5} also gives
$\|\bm\alpha_{zN}^{\circ}\|_2=O_p(N^{-1/2})$ for $z\in\{0,1\}$.
The identity
$\|\bm a+\bm d\|_2^2-\|\bm a\|_2^2
=2\bm a^{\mathsf T}\bm d+\|\bm d\|_2^2$ and Cauchy--Schwarz imply
\begin{align*}
\left|
\|\widehat{\bm\alpha}_{zN}\|_2^2
-
\|\bm\alpha_{zN}^{\circ}\|_2^2
\right|
&\leq
2\|\bm\alpha_{zN}^{\circ}\|_2\|\bm d_{zN}\|_2
+\|\bm d_{zN}\|_2^2\\
&=o_p(N^{-1}),
\qquad z\in\{0,1\},
\end{align*}
and $\|\widehat{\bm\alpha}_{HN}\|_2^2=o_p(N^{-1})$.
The triangle inequality also gives
$\|\widehat{\bm\alpha}_{zN}\|_2
\leq\|\bm\alpha_{zN}^{\circ}\|_2+\|\bm d_{zN}\|_2
=O_p(N^{-1/2})$, and hence
$\|\widehat{\bm\alpha}_{zN}\|_2^2=O_p(N^{-1})$.
The same coupling and terminal-weight condition give
$\|\widehat{\bm\alpha}_{0N}^{C}-\bm\alpha_{0N}^{\circ}\|_2=o_p(N^{-1/2})$,
so the same squared-norm argument applies to the full-length
concurrent-only coefficient used in $\widehat V_C$. Zero-padding preserves
the norm of the arm-compressed vector $\bm a_0^C$.

Lemma~\ref{lem:supp_residual_variance_consistency} gives
$\widehat\sigma_z^2-\sigma_z^2=o_p(1)$ and
$\widehat\sigma_H^2=O_p(1)$. Together with the coefficient bounds, adding and
subtracting
$\sum_{z=0}^1\sigma_z^2\|\widehat{\bm\alpha}_{zN}\|_2^2$ gives
\begin{align*}
\widehat V_{\mathrm{lin}}-V_N^{\circ}
={}&
\sum_{z=0}^1
(\widehat\sigma_z^2-\sigma_z^2)
\|\widehat{\bm\alpha}_{zN}\|_2^2\\
&+
\sum_{z=0}^1\sigma_z^2
\left(
\|\widehat{\bm\alpha}_{zN}\|_2^2
-
\|\bm\alpha_{zN}^{\circ}\|_2^2
\right)
+
\widehat\sigma_H^2\|\widehat{\bm\alpha}_{HN}\|_2^2\\
=&\ o_p(N^{-1}).
\end{align*}
The first sum is $o_p(N^{-1})$ because the two factors in each product have
orders $o_p(1)$ and $O_p(N^{-1})$. The second sum is $o_p(N^{-1})$
by the squared-norm bounds above. The historical
term is $o_p(N^{-1})$ because its two factors have orders $O_p(1)$ and
$o_p(N^{-1})$.

Dividing by the oracle variance gives the required ratios.
The preceding display shows
$N(\widehat V_{\mathrm{lin}}-V_N^{\circ})\xrightarrow{p}0$, whereas
$NV_N^{\circ}\xrightarrow{p}v>0$ by \condref{C5}. Therefore
$\widehat V_{\mathrm{lin}}/V_N^{\circ}-1
=N(\widehat V_{\mathrm{lin}}-V_N^{\circ})/(NV_N^{\circ})
\xrightarrow{p}0$. Replacing $\widehat{\bm\alpha}_{0N}$ by
$\widehat{\bm\alpha}_{0N}^{C}$ in the same calculation gives
$\widehat V_C/V_N^{\circ}\xrightarrow{p}1$. Since the maximum is a continuous
map, $\widehat V_N=\max\{\widehat V_{\mathrm{lin}},\widehat V_C\}$ also satisfies
$\widehat V_N/V_N^{\circ}\xrightarrow{p}1$, proving
\eqref{eq:supp_variance_ratio}.
\end{proof}

\subsection{Proof of Studentized Asymptotic Normality}
\label{supp:studentized_proof}

\begin{proof}[Proof of Theorem~1]
Lemma~\ref{lem:supp_oracle_clt} gives
$(\widehat\Delta_N^{\circ}-\Delta_N-B_N^{\circ})/
\sqrt{V_N^{\circ}}\xrightarrow{d}\Normal(0,1)$.
Lemma~\ref{lem:supp_oracle_equivalence} then gives
$\widehat\Delta_N-\widehat\Delta_N^{\circ}=o_p(N^{-1/2})$.
Because $NV_N^{\circ}\xrightarrow{p}v>0$, it follows that
$(\widehat\Delta_N-\widehat\Delta_N^{\circ})/
\sqrt{V_N^{\circ}}\xrightarrow{p}0$.
Finally, Lemma~\ref{lem:supp_variance_consistency} gives
$\widehat V_N/V_N^{\circ}\xrightarrow{p}1$.

The remaining term is the integrated smoothing bias.
Condition \condref{C5} and the variance consistency established above imply
$N\widehat V_N\xrightarrow{p}v$, so
$1/\sqrt{\widehat V_N}=O_p(\sqrt N)$. Condition \condref{C3}
therefore implies $B_N^{\circ}/\sqrt{\widehat V_N}=o_p(1)$.
The exact decomposition
\begin{align*}
\frac{\widehat\Delta_N-\Delta_N}{
\sqrt{\widehat V_N}}
={}&
\frac{\widehat\Delta_N^{\circ}-\Delta_N-B_N^{\circ}}
{\sqrt{V_N^{\circ}}}
\sqrt{\frac{V_N^{\circ}}{\widehat V_N}}\\
&+
\frac{\widehat\Delta_N-\widehat\Delta_N^{\circ}}
{\sqrt{V_N^{\circ}}}
\sqrt{\frac{V_N^{\circ}}{\widehat V_N}}
+
\frac{B_N^{\circ}}{\sqrt{\widehat V_N}}
\end{align*}
has a first term converging in distribution to a standard normal variable by
Lemma~\ref{lem:supp_oracle_clt} and variance ratio consistency. The second term
vanishes by Lemma~\ref{lem:supp_oracle_equivalence}, and the third vanishes by
the bias bound above. Slutsky's theorem therefore proves the convergence
asserted in Theorem~1.
\end{proof}

\begin{proof}[Proof of Corollary~\ref{cor:radish_wald_validity}]
Let $T_N=(\widehat\Delta_N-\Delta_N)/
\sqrt{\widehat V_N}$. Theorem~1 gives
$T_N\xrightarrow{d}\Normal(0,1)$. Because the standard normal distribution is
continuous at $\pm z_{1-\alpha/2}$,
$\Pr\{-z_{1-\alpha/2}\leq T_N\leq z_{1-\alpha/2}\}\longrightarrow1-\alpha$.
The event in this probability is equivalent to
$|\widehat\Delta_N-\Delta_N|\leq
z_{1-\alpha/2}\sqrt{\widehat V_N}$, which is exactly the event
that the interval in Equation~(8) of the main text contains
$\Delta_N$. This proves the coverage statement. Under $H_0:\Delta_N=0$, the
test rejects precisely on the complementary event $|T_N|>z_{1-\alpha/2}$.
Its probability therefore converges to $\alpha$.
\end{proof}

\section{Algorithms}
\label{supp:algorithms}

\begin{algorithm}[H]
\small
\SetAlgoLined
\SetKwInput{KwIn}{Input}
\SetKwInput{KwOut}{Output}
\KwIn{Historical control data $\mathcal{D}_H = \{(\bm{X}_j^H, Y_j^H)\}_{j=1}^{N_H}$;
      kernel bandwidth $\bm{h}$;
      variance floor $\epsilon_v = 10^{-12}$.}
\KwOut{Local reference center $\theta(\bm{x})$ and undiscounted reference variance
$\hat{\tau}_H^2(\bm{x})$ for each query point $\bm{x}$.}
\BlankLine
\For{each query covariate $\bm{x}$}{
  $w_j^{H}(\bm{x}) \leftarrow K_{\bm{h}}(\bm{X}_j^H - \bm{x})$,\quad $j = 1,\ldots,N_H$\;
  $\theta(\bm{x}) \leftarrow \bigl[\sum_j w_j^{H}(\bm{x}) Y_j^H\bigr] / \bigl[\sum_j w_j^{H}(\bm{x})\bigr]$\;
  $\hat{\sigma}_H^2(\bm{x}) \leftarrow \max\!\left\{\bigl[\sum_j w_j^{H}(\bm{x})\{Y_j^H - \theta(\bm{x})\}^2\bigr] / \bigl[\sum_j w_j^{H}(\bm{x})\bigr],\epsilon_v\right\}$\;
  $n_H^{\mathrm{eff}}(\bm{x}) \leftarrow \bigl[\sum_j w_j^{H}(\bm{x})\bigr]^2 / \bigl[\sum_j \{w_j^{H}(\bm{x})\}^2\bigr]$\;
  $\hat{\tau}_H^2(\bm{x}) \leftarrow \max\{\hat{\sigma}_H^2(\bm{x}) / n_H^{\mathrm{eff}}(\bm{x}),\epsilon_v\}$\;
}
\KwRet{$\{\theta(\bm{x}), \hat{\tau}_H^2(\bm{x})\}$.}
\caption{Stage~I: local historical reference construction.}
\label{alg:stage1}
\end{algorithm}

\begin{algorithm}[H]
\small
\SetAlgoLined
\SetKwInput{KwIn}{Input}
\SetKwInput{KwOut}{Output}
\KwIn{Stage~I outputs $\{\theta(\bm{x}), \hat{\tau}_H^2(\bm{x})\}$;
      planned trial size $N$;
      kernel function $K_{\bm{h}}$ and bandwidth $\bm{h}$;
      safeguards $n_{\mathrm{init}}=20$, $n_{\min}=5$,
      $\epsilon_c=\epsilon_v=10^{-12}$.}
\KwOut{Treatment assignments $\{Z_1, \ldots, Z_N\}$ and allocation-time information ratios $\{R_n^A(\bm X_{n+1})\}_{n=0}^{N-1}$.}
\BlankLine
\For{$n = 0, 1, \ldots, N - 1$}{
  Observe new covariate $\bm{x} \equiv \bm{x}_{n+1}$; evaluate $\theta(\bm{x}), \hat{\tau}_H^2(\bm{x})$ using the Stage~I reference\;
  \uIf{$n<n_{\mathrm{init}}$}{Set $R_n^{A}(\bm x)\leftarrow1$; draw $Z_{n+1}\sim\mathrm{Bernoulli}(1/2)$, observe $Y_{n+1}$, append the record, and continue\;}
  Compute $M_{0,n}(\bm{x})$, $n_{0,n}^{\mathrm{eff}}(\bm{x})$, $\bar{Y}_{0,n}(\bm{x})$, and $\hat{\sigma}_{0,n}^2(\bm{x})$ via $K_{\bm{h}}$\;
  $\widetilde n_{0,n}^{\mathrm{eff}}(\bm{x}) \leftarrow \max\{n_{0,n}^{\mathrm{eff}}(\bm{x}), n_{\min}\}$;\quad
  $\hat\sigma_{0,n}^2(\bm x)\leftarrow\max\{\hat\sigma_{0,n}^2(\bm x),\epsilon_v\}$\;
  $\mathrm{SE}_n^2(\bm{x}) \leftarrow \max\!\left\{\hat{\tau}_H^2(\bm{x}) + \hat{\sigma}_{0,n}^2(\bm{x}) / \widetilde n_{0,n}^{\mathrm{eff}}(\bm{x}),\epsilon_v\right\}$;\quad
  $D_n(\bm{x}) \leftarrow \bigl[-\log\max\{2\Phi(-|\{\bar{Y}_{0,n}(\bm{x})-\theta(\bm{x})\}/\mathrm{SE}_n(\bm{x})|),\epsilon_c\}-1\bigr]_+$;\quad
  $R_n^A(\bm{x}) \leftarrow 1 + \hat\sigma_{0,n}^2(\bm{x})\exp\{-D_n(\bm{x})\}/[\widetilde n_{0,n}^{\mathrm{eff}}(\bm{x})\mathrm{SE}_n^2(\bm{x})]$\;
  Compute the treatment-arm kernel mass $M_{1,n}(\bm{x})$ via $K_{\bm{h}}$\;
  $\phi_n(\bm{x}) \leftarrow \{R_n^A(\bm{x})M_{0,n}(\bm{x})\}^2 \big/ \bigl[\{R_n^A(\bm{x})M_{0,n}(\bm{x})\}^2 + \{M_{1,n}(\bm{x})\}^2\bigr]$\;
  Draw $Z_{n+1} \sim \mathrm{Bernoulli}\{\phi_n(\bm{x})\}$, observe $Y_{n+1}$\;
}
\KwRet{$\{Z_i\}_{i=1}^N$ and $\{R_n^A(\bm X_{n+1})\}_{n=0}^{N-1}$.}
\caption{Stage~II: sequential information-ratio randomization.}
\label{alg:stage2}
\end{algorithm}

\clearpage
The algorithm uses the abbreviations
$D_{N,i}=D_N(\bm X_i)$,
$D^F_{N,i}=D^F_N(\bm X_i)$,
$\mathrm{SE}_{N,i}^2=\mathrm{SE}_N^2(\bm X_i)$,
$R_{N,i}^F=R_N^F(\bm X_i)$, and
$W_{N,i}^F=W_N^F(\bm X_i)$
when evaluating these quantities at $\bm X_i$.

\begin{algorithm}[H]
\small
\SetAlgoLined
\SetKwInput{KwIn}{Input}
\SetKwInput{KwOut}{Output}
\KwIn{Trial data $\{(\bm{X}_i, Y_i, Z_i)\}_{i=1}^N$;
      historical control data $\mathcal{D}_H$;
      Stage~I outputs $\{\theta(\bm{x}),\hat\tau_H^2(\bm{x})\}$;
      kernel function $K_{\bm h}$ with bandwidth $\bm{h}$;
      safeguards $n_{\min}=5$,
      $\epsilon_c=\epsilon_v=10^{-12}$;
      significance level $\alpha=0.05$.}
\KwOut{Point estimate $\widehat{\Delta}_N$, variance estimate
$\widehat{V}_N$, $100(1-\alpha)\%$ Wald interval.}
\BlankLine
Compute the marginal score $D_N^G$ from the completed concurrent and
historical control means and variances\;
\For{$i = 1, \ldots, N$}{
  Form kernel rows $\bm{S}_1(\bm{X}_i), \bm{S}_0(\bm{X}_i), \bm{S}_H(\bm{X}_i)$ and row-normalize\;
  Compute final $n_{0,N}^{\mathrm{eff}}(\bm X_i)$, $\bar Y_{0,N}(\bm X_i)$, and $\hat\sigma_{0,N}^2(\bm X_i)$ from all concurrent controls\;
  $\widetilde n_{0,N}^{\mathrm{eff}}(\bm{X}_i) \leftarrow \max\{n_{0,N}^{\mathrm{eff}}(\bm{X}_i), n_{\min}\}$;\quad
  $\hat\sigma_{0,N}^2(\bm X_i)\leftarrow\max\{\hat\sigma_{0,N}^2(\bm X_i),\epsilon_v\}$\;
  $\mathrm{SE}_{N,i}^2\leftarrow\max\!\left\{\hat\tau_H^2(\bm X_i)+\hat\sigma_{0,N}^2(\bm X_i)/\widetilde{n}_{0,N}^{\mathrm{eff}}(\bm X_i), \epsilon_v\right\}$\;
  $D_{N,i}\leftarrow\bigl[-\log\max\{2\Phi(-|\{\bar Y_{0,N}(\bm X_i)-\theta(\bm X_i)\}/\mathrm{SE}_{N,i}|),\epsilon_c\}-1\bigr]_+$\;
  $D_{N, i}^F \leftarrow D_{N, i}+\{1-\exp[-D_{N, i}]\}D_N^G.$ \;
  $R_{N,i}^F\leftarrow1+\exp(-D^F_{N,i})\hat\sigma_{0,N}^2(\bm X_i)/[\widetilde{n}_{0,N}^{\mathrm{eff}}(\bm X_i)\mathrm{SE}_{N,i}^2]$;\quad
  $W_{N,i}^F\leftarrow(R_{N,i}^F-1)/R_{N,i}^F$\;
  $\widetilde{\bm{S}}_0(\bm{X}_i) \leftarrow \{1 - W_{N,i}^F\} \bm{S}_0(\bm{X}_i)$;\quad
  $\widetilde{\bm{S}}_H(\bm{X}_i) \leftarrow W_{N,i}^F \bm{S}_H(\bm{X}_i)$\;
}
$\bm{a}_1 \leftarrow N^{-1}\sum_i \bm{S}_1(\bm{X}_i)$;\quad
$\bm{a}_0 \leftarrow N^{-1}\sum_i \widetilde{\bm{S}}_0(\bm{X}_i)$;\quad
$\bm{a}_H \leftarrow N^{-1}\sum_i \widetilde{\bm{S}}_H(\bm{X}_i)$\;
$\widehat{\Delta}_N \leftarrow \bm{a}_1^{\top}\bm{Y}_1 - \bm{a}_0^{\top}\bm{Y}_0 - \bm{a}_H^{\top}\bm{Y}_H$\;
For $z=0,1$, set $n_z\leftarrow N_z$; for $z=H$, set
$(n_z,\bm X_z,\bm Y_z)\leftarrow(N_H,\bm X_H,\bm Y_H)$\;
\For{$z \in \{0, 1, H\}$}{
  $\bm{S}_{zz} \leftarrow \operatorname{normRows}[\bm K_{\bm{h}}(\bm{X}_z, \bm{X}_z)]$\;
  $\operatorname{dof}_z \leftarrow \max\{n_z - 2\,\operatorname{tr}(\bm{S}_{zz}) + \operatorname{tr}(\bm{S}_{zz}^{\top}\bm{S}_{zz}),1\}$\;
  $\hat{\sigma}_z^2 \leftarrow \max\{\|\bm{Y}_z - \bm{S}_{zz}\bm{Y}_z\|_2^2 / \operatorname{dof}_z,\epsilon_v\}$\;
}
  $\widehat{V}_{\mathrm{lin}} \leftarrow \hat{\sigma}_1^2\|\bm{a}_1\|_2^2 + \hat{\sigma}_0^2\|\bm{a}_0\|_2^2 + \hat{\sigma}_H^2\|\bm{a}_H\|_2^2$\;
$\bm a_0^C\leftarrow N^{-1}\sum_i\bm S_0(\bm X_i)$;\quad
$\widehat V_C\leftarrow\hat\sigma_1^2\|\bm a_1\|_2^2+
\hat\sigma_0^2\|\bm a_0^C\|_2^2$\;
Floor $\widehat V_{\mathrm{lin}}$ and $\widehat V_C$ at $\epsilon_v$\;
$\widehat V_N\leftarrow\max\{\widehat V_{\mathrm{lin}},\widehat V_C\}$\;
$\widehat{\mathrm{se}}_N\leftarrow\sqrt{\widehat V_N}$\;
\KwRet{$\widehat{\Delta}_N$, $\widehat{V}_N$,
$\bigl[\widehat{\Delta}_N \pm z_{1-\alpha/2}\widehat{\mathrm{se}}_N\bigr]$.}
\caption{Stage~III: Final estimation and inference for the treatment effect $\Delta_N$.}
\label{alg:stage3}
\end{algorithm}

\bibliographystyle{apalike}
\bibliography{references}